\documentclass[aps, pra, superscriptaddress, 10pt, twocolumn]{revtex4-1} 

\usepackage{amsmath}
\usepackage{amsfonts}
\usepackage{amssymb}
\usepackage{graphicx}
\usepackage{braket}
\usepackage{appendix}
\usepackage[caption=false]{subfig}
\usepackage{hyperref}

\newcommand{\eq}[1]{\text{Eq.}~\eqref{#1}}
\newcommand{\fig}[1]{\text{Fig.}~(\ref{#1})}
\newcommand{\sect}[1]{\text{Sec.}~\ref{#1}}
\newcommand{\ie}{\textit{i}.\textit{e}.}

\newcommand{\hH}{\hat{H}}
\newcommand{\hU}{\hat{U}}
\newcommand{\hI}{\hat{\mathbb{I}}}
\newcommand{\hs}{\hat{\sigma}}
\newcommand{\hr}{\hat{\rho}}
\newcommand{\tr}[1]{\text{Tr}\left[#1\right]}

\newcommand{\mauro}[1]{}

\begin{document}

\title{Non-equilibrium thermodynamics of quantum processes assisted by transitionless quantum driving: the role of initial state preparation}

\author{Qiongyuan Wu}
\thanks{Email: qwu03@qub.ac.uk}
\affiliation{Centre for Theoretical Atomic, Molecular and Optical Physics, School of Mathematics and Physics, Queen’s University Belfast, BT7 1NN Belfast, United Kingdom}
\author{Giovanni Barontini}
\affiliation{Midlands Ultracold Atom Research Centre, School of Physics and Astronomy, University of Birmingham, Edgbaston, Birmingham B15 2TT, United Kingdom}
\author{Mauro Paternostro}
\affiliation{Centre for Theoretical Atomic, Molecular and Optical Physics, School of Mathematics and Physics, Queen’s University Belfast, BT7 1NN Belfast, United Kingdom}

\date{\today}

\begin{abstract}
Adiabatic evolution is considered to be the ideal situation for most thermodynamic cycles, as it allows for the achievement of maximum efficiency. However, no power output is produced in light of the infinite amount of time required to perform a transformation. Such issue can be overcome through Shortcuts-To-Adiabaticity (STA) protocols, which allows a dynamics to mimic its adiabatic counterpart, but in a finite time. Transitionless quantum driving (TQD) is one form of STA. We develop and study the effects of TQD on a single-qubit system subjected to a changing magnetic field and a general two-qubit Heisenberg model with a time-dependent interaction strength. We establish a quantitative relation between the work produced across the dynamics and the entropy production resulting from the evolution at hand, focusing on the role played by the initial states of the work medium. We identify the states that extremize performance under the TQD protocol. The thermodynamic implications of STA are discussed in the end.
\end{abstract}  

\maketitle

\section{Introduction}
Quantum thermodynamics extends the framework of classical thermodynamics to the microscopic scale, where the system is subjected to both quantum and thermodynamic fluctuations~\cite{Kosloff_2013, Brandao_2015, Vinjanampathy_2016}. It addresses issues like work and heat when the ensemble size is below the thermodynamic limit \cite{Horodecki_2013}, the properties of systems brought out of equilibrium~\cite{Rio_2011, Hsiang_2018, Santos_2019}, and the efficiency of thermo-engine in the quantum regime~\cite{Kim_2011, Rossnagel_2016, Abah_2017}. As a burgeoning field, quantum thermodynamics holds the promises to impact significantly on quantum technologies~\cite{Goold_2016,Deffner_2019} by providing the energetic footprint of processing information quantum mechanically. Remarkably, some pioneer experimental tests have already been reported~\cite{Liphardt_2002, An_2015, Ara_jo_2018, Klatzow_2019, von_Lindenfels_2019, Peterson_2019}. \mauro{add references by Batalhao,  Obinna's paper on the spin engine, the one by Uzdin, the freewheel by Poshinger} When used to characterize the performance of a thermal machine~\cite{Feldmann_2003, Quan_2007, Quan_2009, Barontini_2019} 
quantum thermodynamics reveals 
that a quantum thermo-engine has the potential to even surpass the theoretic threshold of Carnot bound with the help of quantum resources~\cite{Scully_2002, Dillenschneider_2009, Scully_2011}. 

It is well known that maximum efficiency of a thermodynamic cycle can be reached when the adiabatic condition is fulfilled, namely the work medium evolves quasi-statically. However, such evolution is very time-consuming, resulting in a trade-off between the efficiency of the cycle and the amount of power output. For a quantum work medium, long operating times imply an increase chance of being affected by environmental effects, which would be detrimental to the overall performance of the device. Recently, it has been recognized that very sophisticated quantum control techniques, such as Shortcuts-To-Adiabaticity (STA) protocols~\cite{Berry_2009, Demirplak_2003, Demirplak_2005, Demirplak_2008, GueryOdelin_2019},  can be beneficial in reducing the operating time of quantum cycles, while achieving advantages akin to those of an adiabatic evolution.  In fact, in an STA protocol the Hamiltonian of the work medium is engineered so that the evolution of the system follows an adiabatic trajectory in a finite time. Although many versions of STA protocols have been put forward so far, they can be transformed one into the other~\cite{del_Campo_2013, Campo_2014, Torrontegui_2015}, thus highlighting a common underlying structure. Remarkably, STA protocols have been tested experimentally \cite{Schaff_2011, An_2016} and applied to open quantum dynamics~\cite{Martinez_2016, Dann_2019, Villazon_2019}, which makes them suitable tools for the assessment of quantum thermodynamics. 

In this work, we study the effects of transitionless (counterdiabatic) quantum driving (TQD) techniques on work extraction games applied to both a single- and an interacting two-qubit model. The single-qubit system is subjected to a time-dependent magnetic field, while the coupling strength between the interacting qubits in the second model  can be varied. We construct TQD protocols for both models and 
investigate work and irreversible entropy production across the dynamics when the initial state of the work media is changed and a TQD protocol is applied. We show the existence of a quantitative relation between work and entropy production that depends on the state of the media. We are thus able to identify the states of the latter that extremise their performance under the chosen TQD protocol. Remarkably, we show that the states that maximize the amount of extractable work at a given value of irreversible entropy production are either diagonal or maximally coherent ones. Our results are relevant for a broad range of experimental platforms, such as arrays of ultracold molecules \cite{Anderegg_2019} and Rydberg atoms trapped in optical tweezers \cite{Omran_2019}, in which TQD strategies could boost the performance of quantum information protocols.

\begin{figure*}[tb]
\subfloat[\label{fig:HtqdandFidelityHtqd}]{%
  \includegraphics[width=0.7\columnwidth]{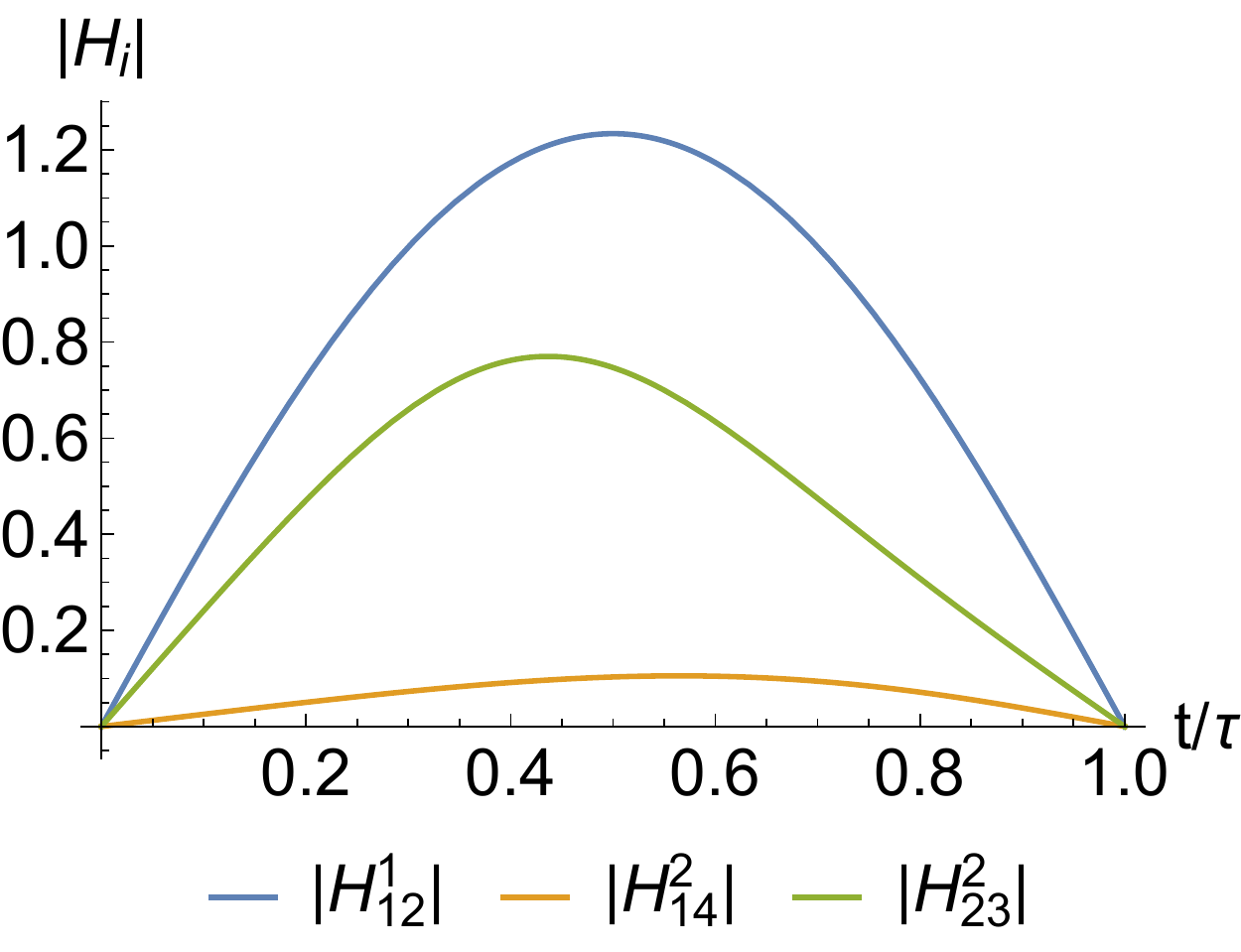}%
}
\hspace{0.3\columnwidth}
\subfloat[\label{fig:HtqdandFidelityFidelity}]{%
  \includegraphics[width=0.7\columnwidth]{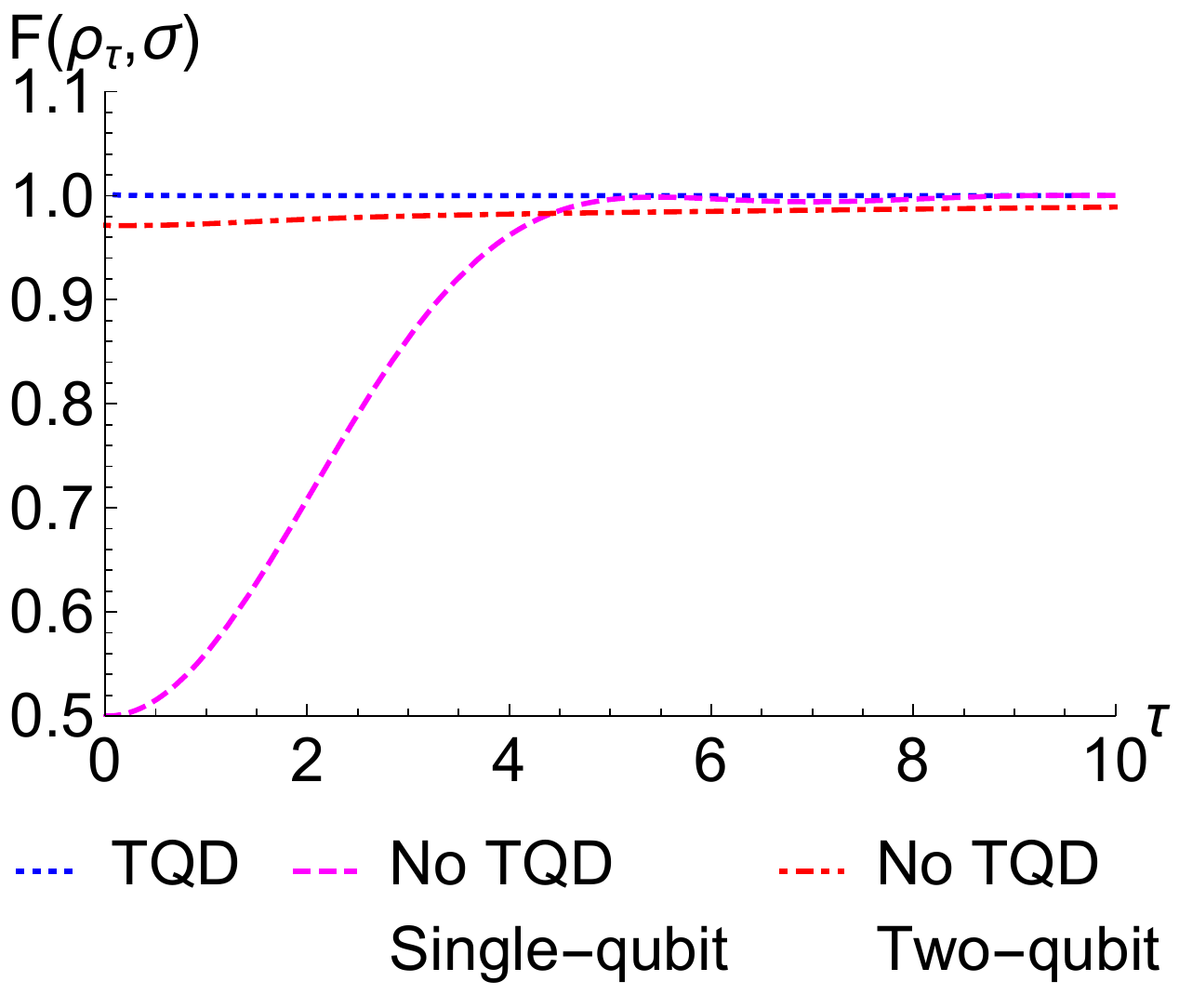}%
}
\caption{Features of the TQD Hamiltonians $\hH_{tqd}(s)$ for both the models. \fig{fig:HtqdandFidelityHtqd} demonstrates the dynamics of derived $\hH_{tqd}$ in Eqs.~(\ref{equ:Htqdonequbit}) and (\ref{equ:Htqdtwoqubit}), where $|H^i_j|$ represents the absolute value of $H_j$ for model $i$. \fig{fig:HtqdandFidelityFidelity} shows the trend followed by the state fidelity against the time interval $\tau$, with and without application of the TQD protocol. The blue line shows the fidelity between the initial and target states for both models when TQD protocol is applied. The purple (orange) curve shows the fidelity for the single-qubit (two-qubit) model without application of the TQD protocol.}
\label{fig:HtqdandFidelity}
\end{figure*}

The paper is structured as follows. In \sect{sec:models} we give a brief introduction on the TQD protocol, and construct protocols for the two models that we address. In \sect{sec:results} we demonstrate the quantitative relation between extractable work and irreversible entropy production when TQD protocols are applied, and discuss its dependence on the system's initial state. 
In \sect{sec:conclusion} we give our conclusions.

\section{TQD on Single- and Two-qubit Models}\label{sec:models}
In this Section, we first give a short introduction on the TQD protocol and then discuss its applications on two specific models, \ie~a single-qubit subject to a changing magnetic field and a two-qubit interacting model with a time-varying interaction strength. 

\subsection{Transitionless quantum driving}\label{sec:tql}
The idea behind the transitionless quantum driving technique (TQD) is to add an additional counterdiabatic Hamiltonian that cancels out the transition among instantaneous eigenstates of the system throughtout the process. For examples, suppose the time-dependent system Hamiltonian is $\hH_s(t)$ with $t\in[0,\tau]$. A suitable counterdiabatic term $\hH_{tqd}(t)$ is added to the system Hamiltonian so that $\hH'(t)=\hH_s(t)+\hH_{tqd}(t)$ with conditions $\hH_{tqd}(0)=\hH_{tqd}(\tau)=0$. Scope of the procedure is for the new Hamiltonian $\hH'(t)$ would stay diagonal in the instantaneous basis $\{\ket{n(t)}\}$ at all times $t$, and the state of the system to stay unchanged, up to a phase factor, in the instantaneous basis, which is the characteristics of an adiabatic evolution. \mauro{Can we move this derivation (highlighted in boldface) to an appendix? it's not crucial and the paper is already long. We can simply say the following: A straightforward derivation of the form of the counteradiabatic term is presented in Appendix A, which leads to} A straightforward derivation of the form of the counterdiabatic term is presented in Appendix \ref{app:derivationHtqd}, which leads to 
\begin{equation}\label{equ:Htqd}
\hH_{tqd}(t) = - \hU(t) \hH'_{nd}(t) \hU^{-1}(t),
\end{equation}
where $\hH'_{nd}$(t) is the non-diagonal part of the total Hamiltonian in the picture defined by the similarity transformation
\begin{equation}
\hU(t)=\sum_n \ket{n(t)}\bra{n(0)}.
\end{equation}
Here $\{\ket{n(t)}\}$ are the instantaneous eigenstates of $\hH_s(t)$. Should $\hH_{tqd}$(t) of \eq{equ:Htqd} satisfy the boundary conditions, the total Hamiltonian $\hH'(t)$ would drive the system adiabatically but in a finite time $\tau$.

\subsection{Single-qubit model}
To demonstrate the effect of TQD, we first consider the Landau-Zener model \cite{Landau_1965, Zener_1932}, where a qubit is subject to a changing magnetic field. The system Hamiltonian has the form (we use $\hbar=1$ throughout this paper)
\begin{equation}\label{equ:hamiltonianonequbit}
\hH(s)=\frac{\omega(s)}{2} \left( B(s) \hs_x + \sqrt{1-B(s)^2} \hs_z \right),
\end{equation}
where $s=t/\tau$ is the normalised time parameter with $t\in[0,\tau]$. We have introduced the frequency $\omega(s)$, while $B(s)\in[0,1]$ is a parameter accounting for the rotation of the external field from the $\hs_z$ to the $\hs_x$ direction of a reference frame. Given $B(s)$, a shortcut $H_{tqd}(s)$ can be designed following the procedure sketched in \sect{sec:tql}. As an example, we set $B(s)=\sin\left[ \frac{\pi}{2} \sin\left(\frac{\pi s}{2}\right)^2 \right]$ and get the counterdiabatic Hamiltonian
\begin{equation}\label{equ:Htqdonequbit}
\begin{split}
\hH_{tqd} (s) &= 
\begin{pmatrix} 
0 & H^1_{12} \\
H^{1\ast}_{12} & 0
\end{pmatrix} \\
&=
\begin{pmatrix} 
0 & -\frac{i \pi^2}{8\tau}\sin(\pi s) \\
\frac{i \pi^2}{8\tau}\sin(\pi s) & 0
\end{pmatrix}.
\end{split}
\end{equation}
Notice that the choice of $\omega(s)$ does not affect the additional Hamiltonian. It can be seen in \fig{fig:HtqdandFidelityHtqd} that $\hH_{tqd} (s)$ satisfies the boundary conditions $\hH_{tqd}(0)=\hH_{tqd}(1)=0$, while the operation length $\tau$ governs the intensity of the additional Hamiltonian, \ie~small $\tau$ correspond to a sharp change of $|H^1_{12}|$. 

Given this, the total Hamiltonian $\hH'(s)=\hH_s(s)+\hH_{tqd}(s)$ drives the quantum system adiabatically for arbitrary $\tau$. To demonstrate it, we prepare the initial system in state $\ket{\uparrow}$, where we have introduced the eigenstates $\ket{\uparrow}$ and $\ket{\downarrow}$ of $\hs_z$ with eigenvalues $1$ and $-1$, respectively. Should the system have an adiabatic evolution, the final system state would be in state $\ket{+}$, where 
$\hs_x\ket{\pm}=\pm\ket{\pm}$. \mauro{Fidelity is never introduced} 
The effectiveness of this protocol can be measured through the fidelity $F(\rho,\sigma)$ between the evolved state $\rho$ and the targeted state $\sigma$, such that
\begin{equation}
F(\rho,\sigma)=\left(\tr{\sqrt{\sqrt{\rho}\sigma\sqrt{\rho}}}\right)^2,
\end{equation}
which gives the ``closeness'' of two states. The result is shown in \fig{fig:HtqdandFidelityFidelity}, where we set the initial and final states to be $\ket{\uparrow}$ and $\ket{+}$. One can see that, without TQD protocol, the fidelity approaches to $1$ if the time interval $\tau$ is large, according with the adiabatic condition. However, the fidelity can reach $1$ for arbitrary $\tau$ when the TQD protocol is applied.

\subsection{Two-qubit interacting model}
Here we consider the case where the working medium consists of two interacting qubits. The system is described by the Heisenberg Hamiltonian \cite{Takhtadzhan_1979}
\begin{equation}\label{equ:HeisenbergHam}
\begin{split}
\hH_s(s)&= \epsilon_1 \hs_1^z +\epsilon_2 \hs_2^z \\
&+ \frac{J(s)}{2} (\alpha\hs_1^x\hs_2^x+\beta\hs_1^y\hs_2^y+\gamma\hs_1^z\hs_2^z).
\end{split}
\end{equation}
Here $\epsilon_i$ is the frequency of $i^\text{th}$ qubit, $J(s)$ is the overall control parameter of the interaction, and $\{\alpha,\beta,\gamma\}$ are anisotropy parameters. 
We consider the case where only the overall coupling term $J(s)$ is time-dependent. 

As in the single-qubit case, the coupling strength $J(s)$ determines the specific form of counterdiabatic term $\hH_{tqd}(s)$. In order to meet the boundary conditions $\hH_{tqd}(0)=\hH_{tqd}(1)=0$, we choose $J(s)=\cos\left[\frac{\pi}{2}\cos\left( \frac{\pi (s)}{2} \right)\right]$. Given this choice, we generate the counterdiabatic Hamiltonian as
\begin{equation}\label{equ:Htqdtwoqubit}
\hH_{tqd}(s)=\begin{pmatrix} 
0 & 0 & 0 & H^2_{14} \\
0 & 0 & H^2_{23} & 0 \\
0 & H^{2\ast}_{23} & 0 & 0 \\
H^{2\ast}_{14} & 0 & 0 & 0 
\end{pmatrix}
\end{equation}
with 
\begin{widetext}
\begin{equation}\label{equ:H14}
H^2_{14}=-\frac{i\pi^2(\alpha-\beta)\sin{\left(\frac{\pi s}{2}\right)}\sin{\left[\frac{1}{2}\cos{\left(\frac{\pi s}{2}\right)}\right]}(\epsilon_1+\epsilon_2)}{2\tau\left[(\alpha-\beta)^2\left(1+\cos{\left[\pi\cos{\left(\frac{\pi s}{2}\right)}\right]}\right)+8\epsilon_1^2+16\epsilon_1\epsilon_2+8\epsilon_2^2\right]},
\end{equation}
and
\begin{equation}\label{equ:H23}
H^2_{23}=-\frac{i\pi^2(\alpha+\beta)\sin{\left(\frac{\pi s}{2}\right)}\sin{\left[\frac{1}{2}\cos{\left(\frac{\pi s}{2}\right)}\right]}(\epsilon_1-\epsilon_2)}{2\tau\left[(\alpha+\beta)^2\left(1+\cos{\left[\pi\cos{\left(\frac{\pi s}{2}\right)}\right]}\right)+8\epsilon_1^2-16\epsilon_1\epsilon_2+8\epsilon_2^2\right]}.
\end{equation}
\end{widetext}
Such $\hH_{tqd}$(s) satisfies the boundary conditions for all choices of $\{\epsilon_1,\epsilon_2,\alpha,\beta\}$. One specific example can be seen in \fig{fig:HtqdandFidelityHtqd}, where we set $\epsilon_1=1.4, \epsilon_2=2, \alpha=0.3, \beta=1.2$. One notices that $\gamma$ does not occur in Eqs.~(\ref{equ:H14}) and (\ref{equ:H23}). This is because $\hs^1_z\hs^2_z$ does not contribute to the transition during the dynamics since such term commutes with the time-dependent Hamiltonian at all time, $[\hs_z^1\hs_z^2,\hH(s)]=0$. The effectiveness of such $\hH_{tqd}$ can be seen in \fig{fig:HtqdandFidelityFidelity}, where we set the initial state $\hr_0$ and the target state $\sigma$ to be the thermal state in the initial and final bases, respectively. One can see that, without TQD protocol, the fidelity approaches $1$ asymptotically, while such optimal value is reached for any value of $\tau$ when the TQD protocol is applied.

\section{Work and Entropy Production under TQD Protocol}\label{sec:results}
Here we study the thermodynamic consequences of the use of TQD by using the models introduced above as benchmarks. Specifically, we are interested in the work and irreversible entropy production during the dynamics, when different initial states are prepared. The irreversible entropy production can be calculated as
\begin{equation}\label{equ:relativeentropy}
S_{irr}\equiv S(\hr_\tau||\hr_{th})=\tr{\hr_\tau \ln \hr_\tau - \hr_\tau\ln\hr_{th}},
\end{equation}
which is the relative entropy distance between states $\hr_\tau$ and $\hr_{th}$ \cite{Deffner_2010} \mauro{we did not introduce the thermal state $\hr_{th}$}. Here $\hr_\tau$ is the final state of the system after the evolution, $\hr_{th}\equiv\frac{e^{-\beta \hH_\tau}}{Z(\beta)}$ is the hypothetical thermal state for the final system, where $Z(\beta)$ is the partition function with inverse temperature $\beta$ (we set $\beta=1$). On the other hand, the work extracted is calculated as the energy difference between the initial and the final systems,
\begin{equation}\label{equ:workproduction}
\Delta W\equiv\tr{\hr_\tau \hH_\tau}-\tr{\hr_0 \hH_0}.
\end{equation}
For both models, we establish the quantitive relation between work and entropy production when the work medium takes arbitrary initial states. We also calculate the quantative relation between the produced work (entropy) and the amount of coherence contained in the initial state, where the relative entropy of coherence is used as a figure of merit \cite{Baumgratz_2014}. The latter is defined as
\begin{equation}\label{equ:relativecoherence}
C_{rel}(\hr)=E_N(\hr_{diag})-E_N(\hr).
\end{equation}
Here $\hr_{diag}$ is attained by removing all coherence from $\hr$, and $E_N(\cdot)$ is the von Neumann entropy of the state. This measures the relative entropy distance between state $\hr$ and its diagonal counterpart $\hr_{diag}$.

\subsection{Analytical study}
Using the fact that the states remain unchanged -- up to a phase factor -- in the instantaneous basis under the TQD protocol, we can derive the analytical expressions of work and irreversible entropy production for arbitrary initial states. Due to the unchanged populations of different energy levels, the work production is simply the sum of the energy differences of each energy level multiplied by the corresponding population. Namely
\begin{equation}\label{equ:workdependence}
	\begin{split}
		\Delta W =& \tr{\hr_\tau \hH_\tau}-\tr{\hr_0 \hH_0} \\
		=& \sum_{i=1}^n \rho_0^i (E_\tau^i - E_0^i).
	\end{split}
\end{equation}
Here $\rho_0^i$ is the population of the $i^\text{th}$ energy level, while $E_\tau^i$ and $E_0^i$ are the corresponding eigenvalues for the final and initial Hamiltonian, respectively. 

On the other hand, the irreversible entropy production of the process is characterized by the relative entropy between the final state $\hr_\tau$ and the hypothetical thermal state $\hr_{th}$. Here we call $\hr_{th}=\text{diag}(\rho_{th}^1,\dots,\rho_{th}^{n})$, written in the final basis. Due to the unitary evolution, the von Neumann entropy of the final state $\hr_\tau$ is invariant and equals that of the initial state $\hr_0$. Calculating the second term in the basis of the final Hamiltonian, the relative entropy can be written as
\begin{equation}\label{equ:entropydependence}
	\begin{split}
		S_{irr}=&\tr{\hr_\tau\ln\hr_\tau}-\tr{\hr_\tau\ln\hr_{th}} \\
		=& -E_N(\hr_0) - \sum_{i=1}^n\rho_0^i\ln\rho_{th}^i.
	\end{split}
\end{equation}
Both work and irreversible entropy productions depend on the population distribution of the initial state. However, the coherence of the initial state does not contribute to the work production. Instead, it contributes negatively to the irreversible entropy production. In the following, we establish a quantitative relation between work and irreversible entropy production for the two models discussed above. Our approach blends theoretical predictions and numerical validations.

\begin{figure}[tb]
\includegraphics[width=0.7\columnwidth]{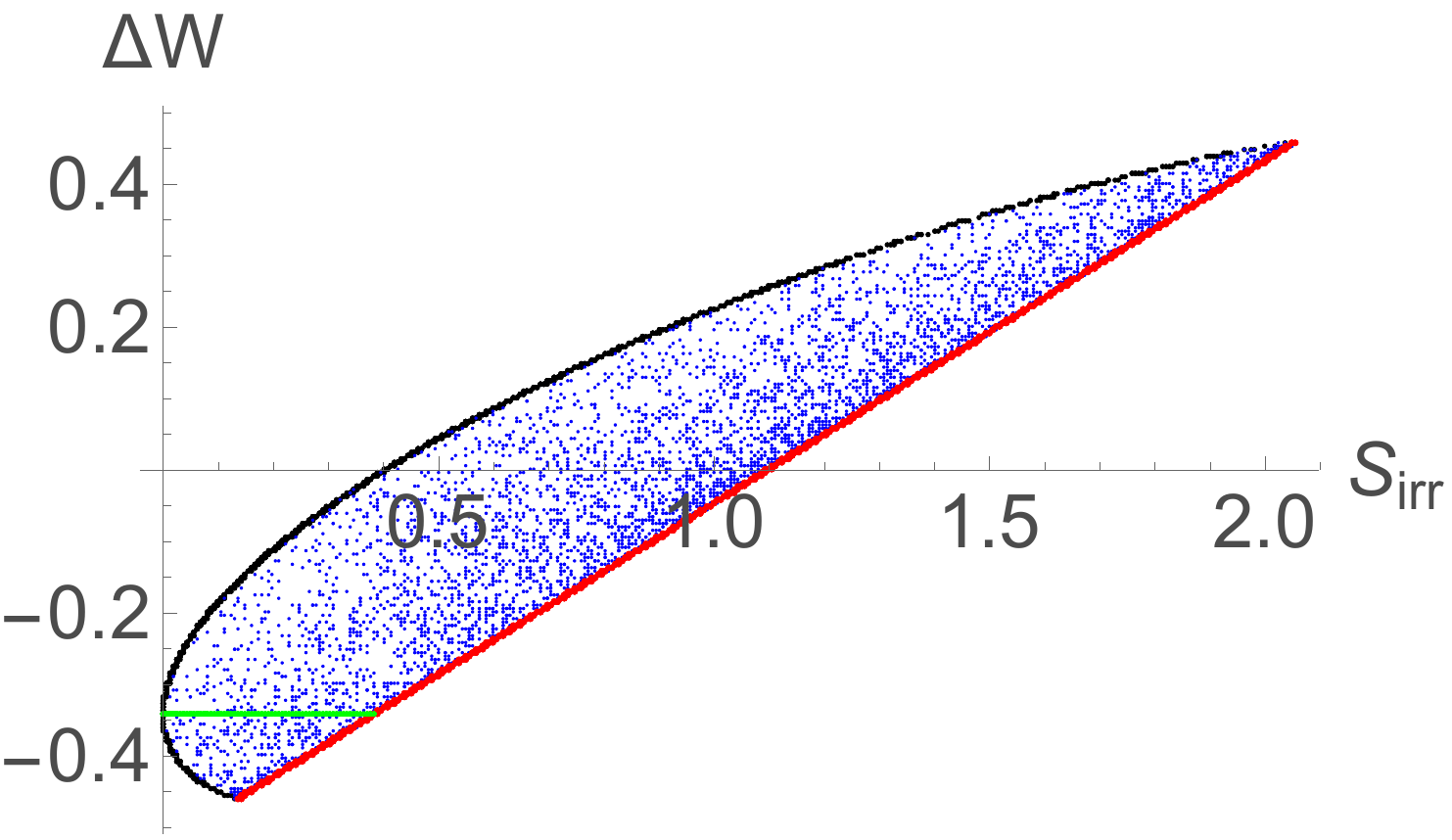}
\caption{Relation between work and irreversible entropy production for the single-qubit model. Here, the blue dots show the results gathered using randomly generated states. The black (red) curve are boundaries provided by diagonal (maximally coherent) states. The green line shows the results achieved using thermal states of the final Hamiltonian, with coherence growing from zero to the maximum value allowed by the model.}
\label{fig:workentropyonequbit}
\end{figure}

\subsection{Single-qubit model}
Let us consider the system Hamiltonian given in \eq{equ:hamiltonianonequbit} with the shortcut in \eq{equ:Htqdonequbit}. As $\omega(s)$ does not contribute to the shortcut Hamiltonian, we set $\omega(s)=1+\tanh{\left(\frac{\pi s}{2}\right)}$ without loss of generality. 

\begin{figure*}[tb]

\subfloat[\label{fig:workcoherenceonequbit}]{%
  \includegraphics[width=0.7\columnwidth]{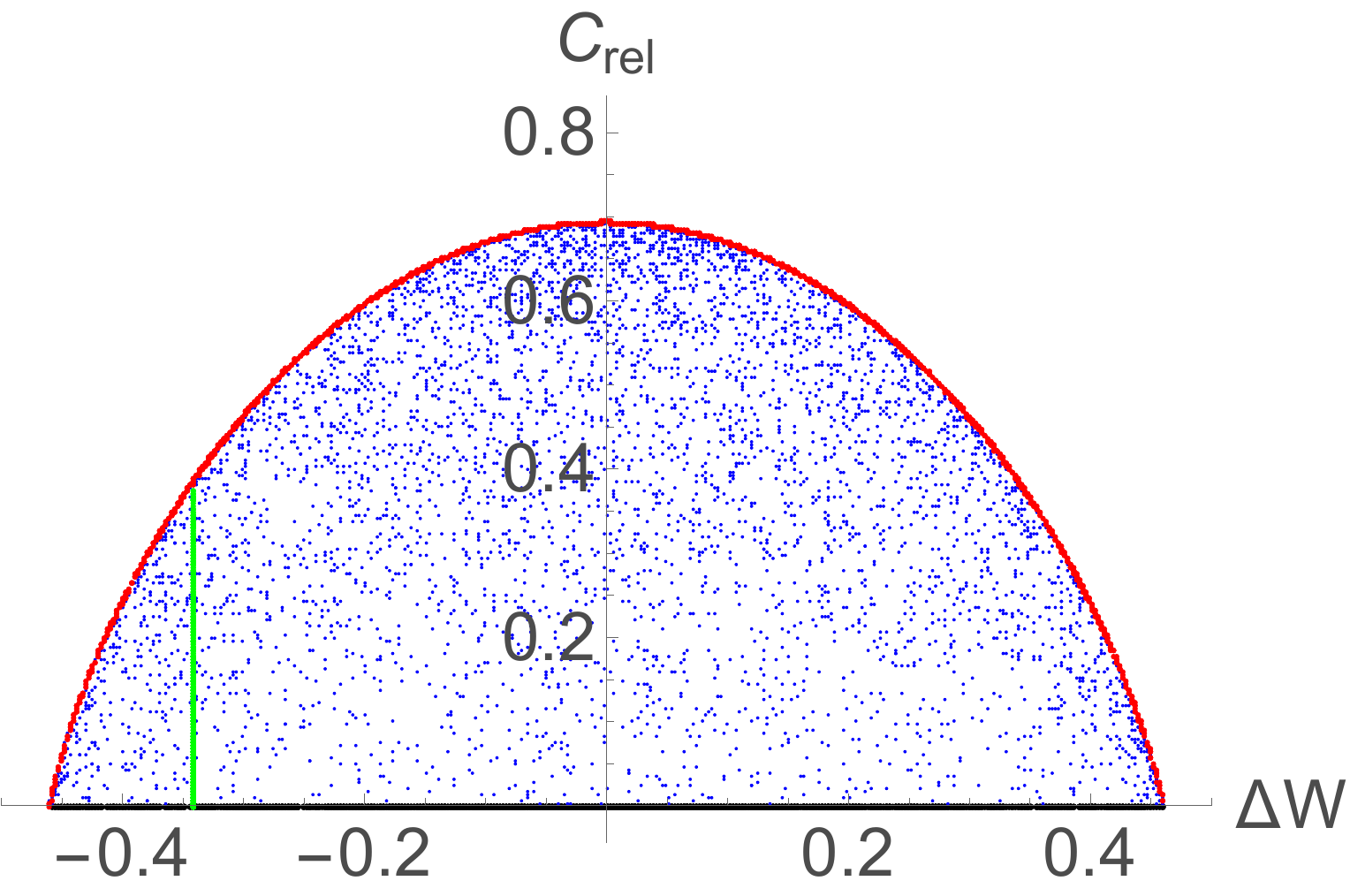}%
}
\hspace{0.3\columnwidth}
\subfloat[\label{fig:entropycoherenceonequbit}]{%
  \includegraphics[width=0.7\columnwidth]{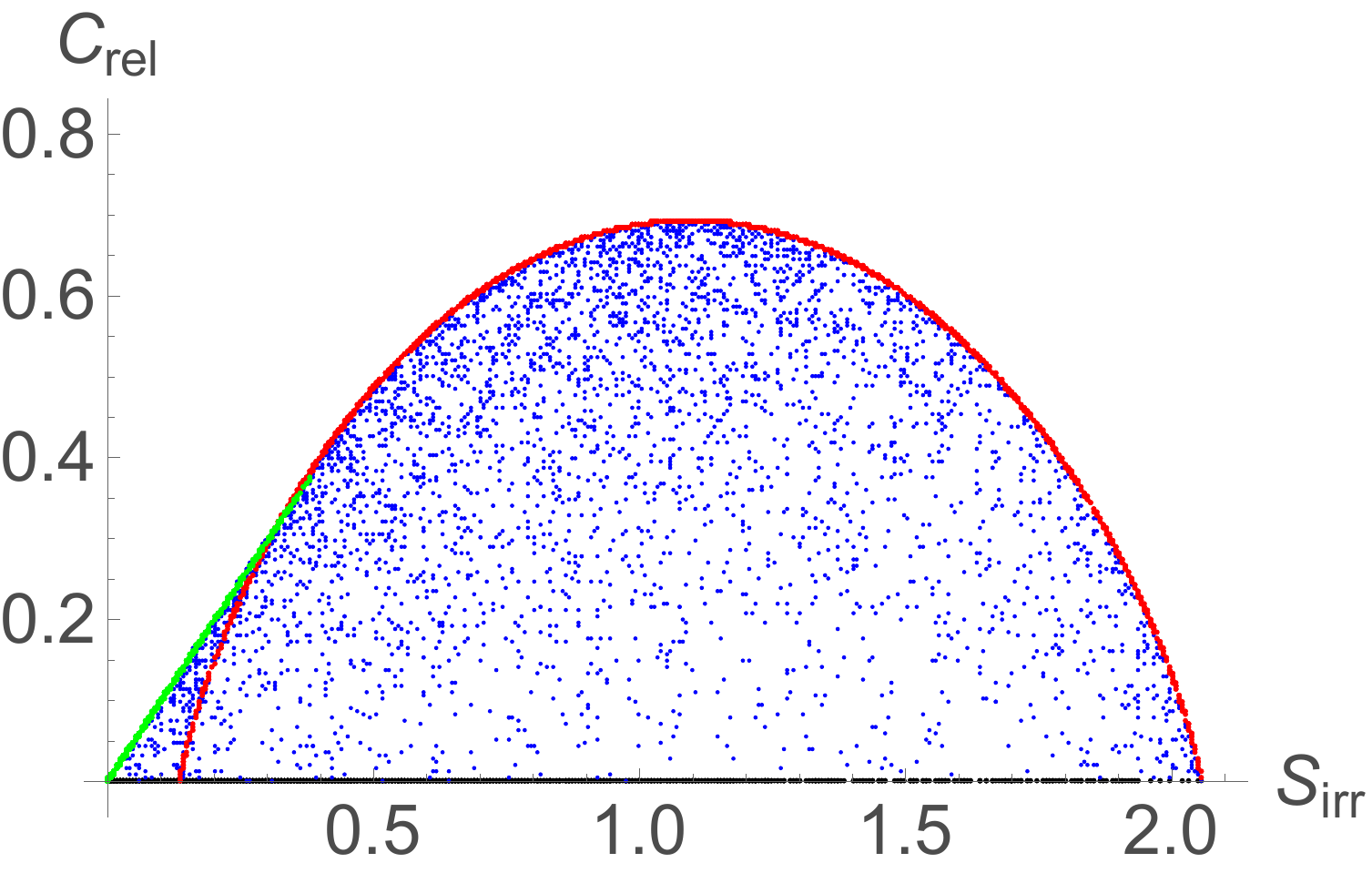}%
}

\caption{\small Relation between work (irreversible entropy) production and the amount of coherence contained in the initial state. For both panels, the blue dots represent randomly generated states, the black line represents diagonal states, while the red curve represents maximally coherent states. The green line represent the results achieved using the thermal states of the final Hamiltonian, with growing coherences. Such states form part of the upper boundary in panel {\bf (b)}.}
\label{fig:workentropycoherenceonequbit}
\end{figure*}


Analytical expressions for work and irreversible entropy production can be attained from Eqs.~(\ref{equ:workdependence}) and (\ref{equ:entropydependence}). Given the initial state 
\begin{equation}
\hr_{in}=\begin{pmatrix}
a_{in} & c_{in}  \\
c_{in}^\ast & 1-a_{in} \\
\end{pmatrix},
\end{equation}
the work extracted under the TQD protocol reads 
\begin{equation}
\Delta W = 
\left(\frac{1}{2}-a_{in}\right)\tanh\left(\frac{\pi}{2}\right),
\end{equation}
which depends on the population of the excited state in $\hr_{in}$. On the other hand, the irreversible entropy production is
\begin{equation}
\begin{split}
S_{irr} &= -E_N(\hr_{in}) - a_{in} \ln\rho_{th}^{1}-(1-a_{in})\ln\rho_{th}^{2} \\
&= -E_N(\hr_{in}) - a_{in} \ln (\rho_{th}^2/\rho_{th}^1) - \ln\rho_{th}^2,
\end{split}
\end{equation}
where $\hr_{th}=\text{diag}(\rho_{th}^1,\rho_{th}^2)$. Here $E_N(\hr_{in})$ is a nonlinear function of entries $a_{in}$ and $|c_{in}|$. However, if the initial state is maximally coherent (pure), which results in $E_N(\hr_{in})=0$, the $S_{irr}$ retrieves the linear dependence on the excited state population $a_{in}$. 

In order to validate the above result, we have generated $5\times10^3$ random initial states (uniformly according to the Haar measure), and calculated work and irreversible entropy production corresponding to such choices. The results are shown in \fig{fig:workentropyonequbit}. As the results gathered in this way match those achieved analytically, we believe the size of such sample to be sufficient. In \fig{fig:workentropyonequbit} we have  highlighted three curves, which represent outcomes for three groups of initial states
\begin{equation}
\begin{aligned}
&\rho_{d}=\begin{pmatrix} 
a & 0  \\
0 & 1-a \\
\end{pmatrix}, \quad \rho_{mc}=\begin{pmatrix} 
a & c  \\
c^\ast & 1-a \\
\end{pmatrix} \\[2ex]
&\text{and} \quad \rho_{cth}=\begin{pmatrix} 
T_1 & c'  \\
c'^\ast & T_2 \\
\end{pmatrix}.
\end{aligned}
\end{equation}
Here $a\in[0,1]$, with $c=\sqrt{a-a^2}$ and $c'\in[0,c]$. $T_1=1-T_2=\frac{\exp{(-\omega_1/2)}}{\exp{(-\omega_1/2)}+\exp{(-\omega_2/2)}}$, with $\omega_1=-\omega_2=\omega(1)$. Specifically, $\rho_d$ is a diagonal state, $\rho_{mc}$ is a maximally coherent (pure) state, and we denote $\rho_{cth}$ as coherent thermal states, which are thermal states in final basis with some random coherence.

In \fig{fig:workentropyonequbit}, the black curve represents diagonal states $\hr_d$ as initial states, red curve represents maximally coherent states $\hr_{mc}$ as initial states, and green line represents the coherent thermal states $\hr_{cth}$ as initial states, respectively. Zero irreversible entropy production can be reached when the initial state is the coherent thermal state with $c'=0$, where the corresponding extracted work is $\Delta W =-0.342102$. The maximum work is extracted when the initial state is a pure excited state, with $\Delta W= -\frac{1}{2}\tanh{(\frac{\pi}{2})}$. Furthermore, the relation between work (irreversible entropy) production and the amount of coherence is shown in \fig{fig:workentropycoherenceonequbit}. Trivially, the maximally coherent states form the upper boundary and the diagonal states form the lower boundary. An interesting feature can be seen in \fig{fig:entropycoherenceonequbit}, where the coherent thermal states have the ability to contain more coherence while producing the same amount of irreversible entropy.

Combining numerical and analytical results, one can draw the following conclusion. For the single-qubit model with TQD protocol, when taking maximally coherent states as initial states, the work production and the irreversible entropy production are linearly correlated, due to the fact that they are linearly dependent on the excited state population $a_{in}$. However, coherence in the initial state does not contribute to the work production, but negatively contribute to the irreversible entropy production. Moreover, coherence and population distributions play different roles in the irreversible entropy production. 

\subsection{Two-qubit interacting model}
In the following we establish the quantitive relation between work and entropy production for two-qubit interacting model, when the work medium takes all possible initial states. The system Hamiltonian is given in \eq{equ:HeisenbergHam} with the shortcut in \eq{equ:Htqdtwoqubit}. 


Analytical expressions of work and irreversible entropy productions can be derived from Eqs.~(\ref{equ:workdependence}) and (\ref{equ:entropydependence}). Suppose the initial state is 
\begin{equation}
\hr_{in}=\begin{pmatrix}
d_1 & c_{12} & c_{13} & c_{14} \\
c_{12}^\ast & d_2 & c_{23} & c_{24} \\
c_{13}^\ast & c_{23}^\ast & d_3 & c_{34} \\
c_{14}^\ast & c_{24}^\ast & c_{34}^\ast & d_4 \\
\end{pmatrix},
\end{equation}
that is positive, semi-definitive and trace-one. The produced work can be expressed as 
\begin{equation}\label{equ:twoqubitwork}
\Delta W = \sum_{i=1}^4 d_i (E_\tau^i - E_0^i),
\end{equation}
where the instantaneous eigenvalues of the system Hamiltonian, given in \eq{equ:HeisenbergHam}, are
\begin{widetext}
\begin{alignat}{2}\label{equ:tdeigenvalues}
&E_s^1=\frac{1}{2}\left( \gamma J_s + \sqrt{(\alpha - \beta)^2 J_s^2+4(\epsilon_1+\epsilon_2)^2} \right), \quad &&E_s^2=\frac{1}{2}\left( -\gamma J_s + \sqrt{(\alpha + \beta)^2J_s^2+4(\epsilon_1-\epsilon_2)^2} \right), \notag\\[2ex] 
&E_s^3=\frac{1}{2}\left( -\gamma J_s - \sqrt{(\alpha + \beta)^2J_s^2+4(\epsilon_1-\epsilon_2)^2} \right), \quad &&E_s^4=\frac{1}{2}\left( \gamma J_s - \sqrt{(\alpha - \beta)^2J_s^2+4(\epsilon_1+\epsilon_2)^2} \right). 
\end{alignat}
\end{widetext}
with $J_0=0$ and $J_\tau=1$. Above eigenvalues illustrate effects of model parameters $\{\epsilon_1,\epsilon_2,\alpha,\beta,\gamma\}$ to the work production of the process. On the other hand, the irreversible entropy production is given as 
\begin{equation}\label{equ:twoqubitentropy}
		S_{irr}= -E_N(\hr_{in}) - \sum_{i=1}^4 d_i \ln\rho_{th}^i.
\end{equation}
where thermal state is $\hr_{th}=diag(\rho_{th}^1,\rho_{th}^2,\rho_{th}^3,\rho_{th}^4)$ in the final basis. Same as in the previous case, coherences in the initial state $\hr_{in}$ do not contribute the work production, and it contributes negatively to the irreversible entropy production. Using these results we can find the boundaries in the work and irreversible entropy relation, as shown below.


To illustrate the relation, we consider a specific isotropic regime such that
\begin{equation}
\hH_s(s)= \hs_1^z + \hs_2^z + \frac{J(s)}{2} (\hs_1^x\hs_2^x+\hs_1^y\hs_2^y).
\end{equation}
We would like to establish the quantitive relation between work and irreversible entropy production with respect to different initial states. Since coherence only contributes negatively to the irreversible entropy production, we consider the most prominent initial states to be diagonal states and maximally coherent (pure) states. For each kind, we generate $10^4$ states randomly, and calculate the corresponding work and irreversible entropy productions.  

\begin{figure}[bb]
\includegraphics[width=0.7\columnwidth]{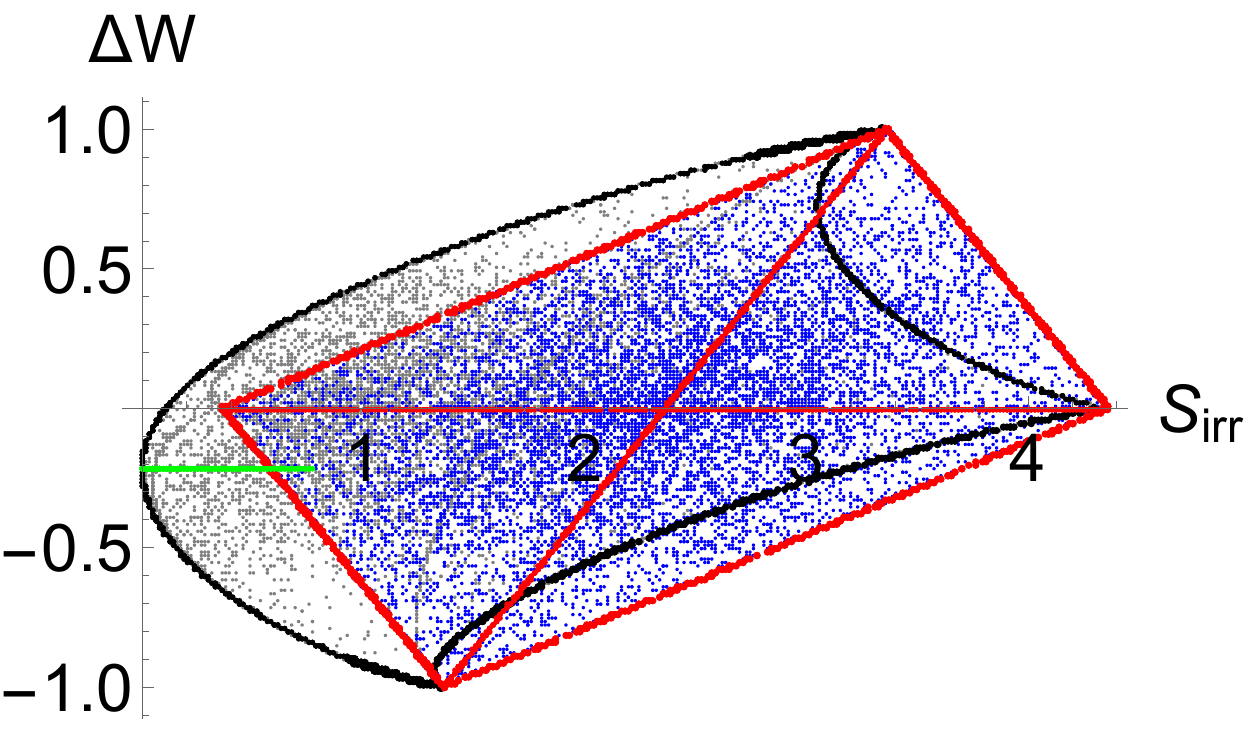}
\caption{Relation between produced work and irreversible entropy production with respect to randomly generated initial states. The blue (grey) dots stand for the results achieved using maximally coherent and pure states (diagonal states). The boundary for diagonal states is shown in black, while the boundary for maximally coherent states is in red. Green dots show the results achieved using thermal states that contain random amount of coherence. The results corresponding to randomly chose initial states are shown as dots within the region obtained by joining such boundaries.}
\label{fig:workandentropyallstates}
\end{figure}

The quantitive relation is shown in \fig{fig:workandentropyallstates}. Here grey dots represent diagonal states with zero coherence, while blue dots represent maximally coherent states. One can observe that, although these two kinds of states partly overlap with each other in the relation, states that produce smallest irreversible entropy are all diagonal states, while states that produce largest irreversible entropy are all maximally coherent states. Boundaries for diagonal and maximally coherent states in the relation can be numerically found given the analytic expressions in Eqs.~(\ref{equ:twoqubitwork}) and (\ref{equ:twoqubitentropy}). The area within the region confined by the black curves represents all possible outcomes of the process when the initial states are diagonal. They are found through the following optimisation problem
\begin{equation}
\begin{aligned}
{\min (\text{or} \max)} \quad & S(\hr_f||\hr_{th}) \\
\textrm{s.t.} \quad &  \Delta W = C\\
  & \hr_{in}=diag(d_1,d_2,d_3,d_4)    \\
\end{aligned}
\end{equation}
where $C\geq0$ is the amount of work assigned manually, $\{d_1,d_2,d_3,d_4\}$ are diagonal terms in the density matrix that need to be found. On the other hand, all possible outcomes for maximally coherent states are confined in the red box. Interestingly, here all six red lines are made up of maximally coherent states that have only two non-zero diagonal terms. For examples, these following two types of states
\begin{equation}
\begin{aligned}
&\rho_{red}^1=\begin{pmatrix} 
a & 0 & c & 0 \\
0 & 0 & 0 & 0 \\
c^\dagger & 0 & 1-a & 0 \\
0 & 0 & 0 & 0
\end{pmatrix}, \\[2ex]
&\rho_{red}^2=\begin{pmatrix} 
a & c & 0 & 0 \\
c^\dagger & 1-a & 0 & 0 \\
0 & 0 & 0 & 0 \\
0 & 0 & 0 & 0
\end{pmatrix}
\end{aligned}
\end{equation}
comprise two right boundaries in \fig{fig:workandentropyallstates}. Here $a\in[0,1]$, and $c=\sqrt{a-a^2}$ is the maximal coherence that can be reached. Numerically we have found that outcomes given by other states all lie in the joined regions in the figure. In this regime, the maximal work extraction can be achieved, such that $\Delta W=-1$, when the initial state is $\hr_{W}=diag(0,0,1,0)$, while the minimal irreversible entropy production can be achieved, such that $S_{irr}=0$, when the initial state is $\hr_{S}=diag(\rho_{th}^1,\rho_{th}^2,\rho_{th}^3,\rho_{th}^4)$. 

\begin{figure*}[tb]
\subfloat[\label{fig:coherenceandwork}]{%
  \includegraphics[width=0.7\columnwidth]{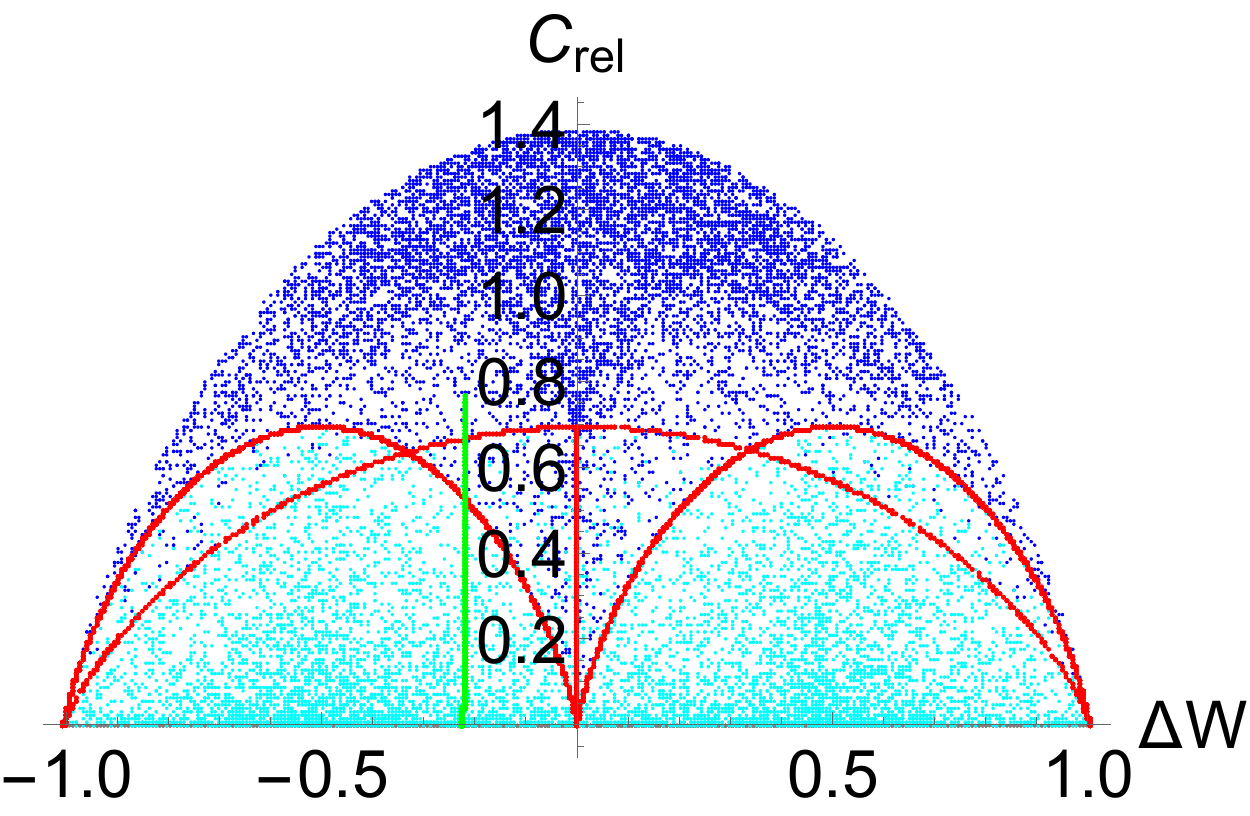}%
}
\hspace{0.3\columnwidth}
\subfloat[\label{fig:coherenceandentropy}]{%
  \includegraphics[width=0.7\columnwidth]{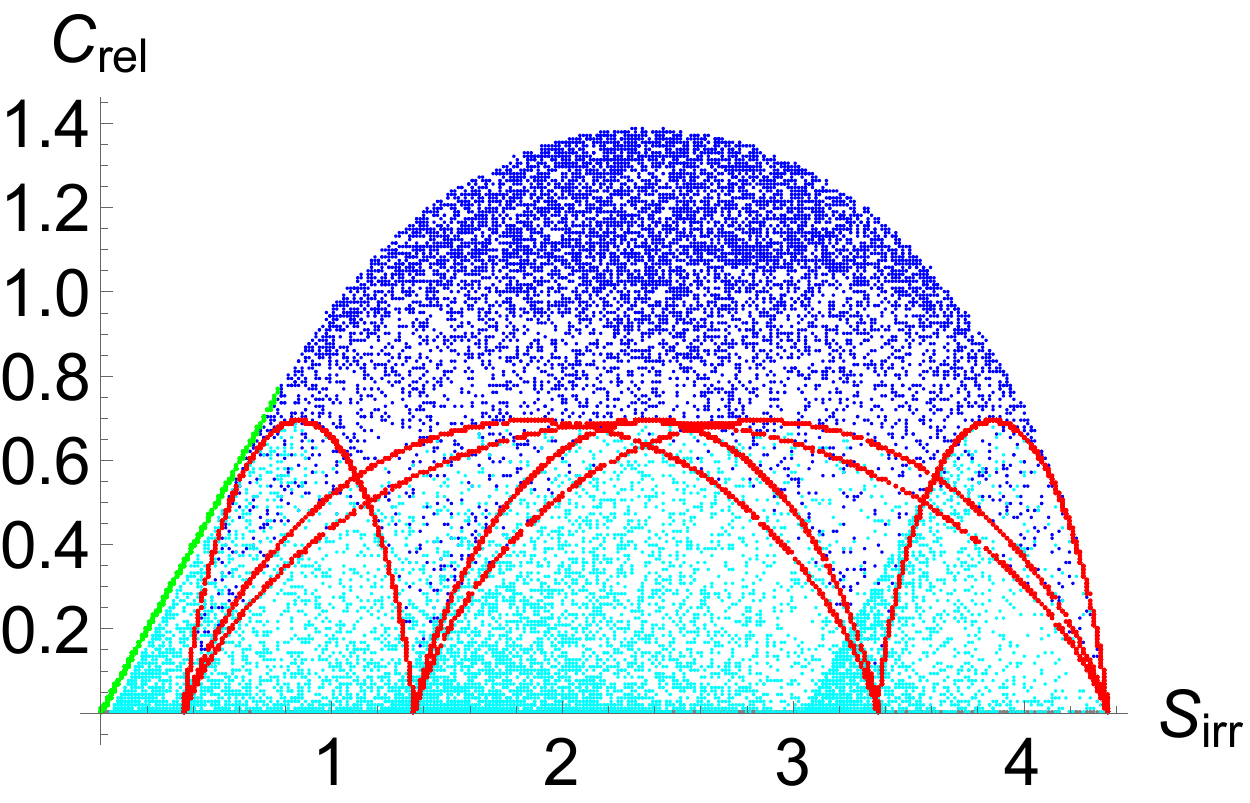}%
}

\caption{Relation between coherence and work (irreversible entropy) production with respect to randomly generated states. Here blue dots represent maximally coherent (pure) states. Red dots represent boundary states in \fig{fig:workandentropyallstates}. Cyan states are boundary states with random amount of coherence. Green dots represent thermal states that contain random amount of coherence.}
\label{fig:workentropycoherence}
\end{figure*}

\begin{figure*}[tb]

\subfloat[\label{fig:workentropyalphagamma}]{%
  \includegraphics[width=0.7\columnwidth]{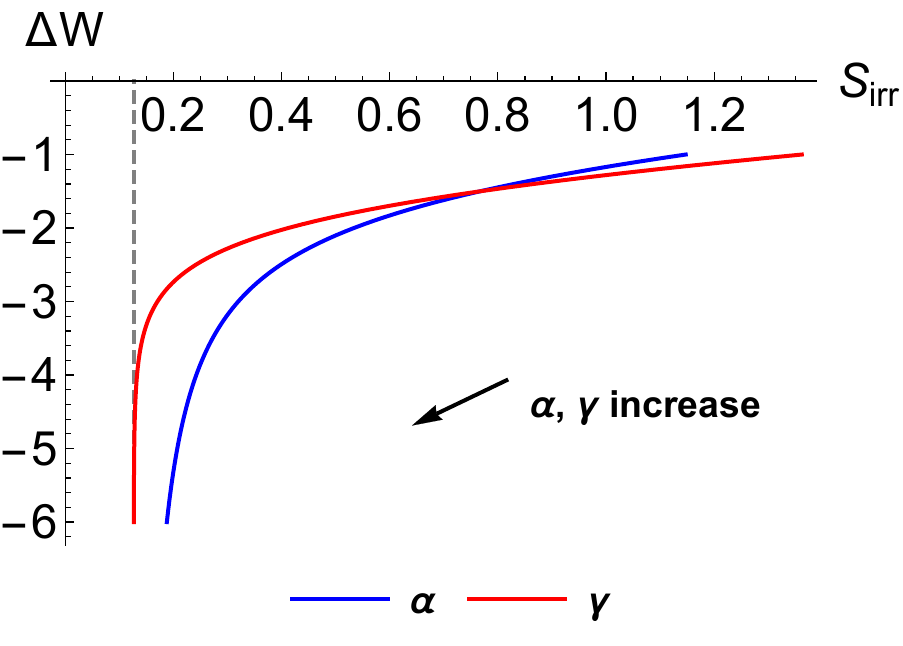}%
}
\hspace{0.3\columnwidth}
\subfloat[\label{fig:workalphagamma}]{%
  \includegraphics[width=0.7\columnwidth]{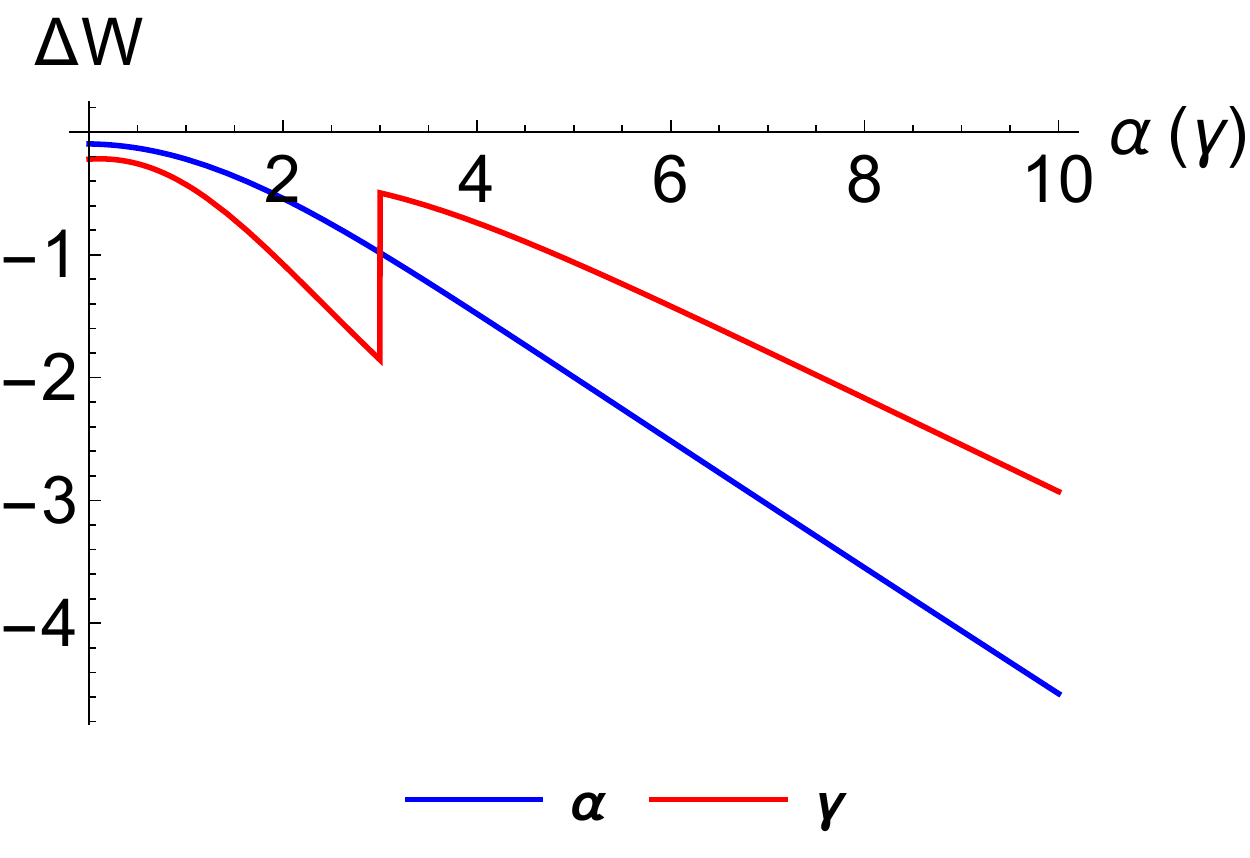}%
}

\caption{ Maximum work production with respect to different regimes. In panel {\bf (a)}, the maximum work extraction diverges as $\alpha$ (or $\gamma$) increases, while the irreversible entropy hits the limit $-\ln\left(\frac{1}{1+e^{-2}}\right)$. In panel~{\bf (b)} the jump of the red curve is caused by an energy crossing in the spectrum of the Hamiltonian of the model being considered.}
\label{fig:maximumworkproduction}
\end{figure*}

The relation between work (irreversible entropy) and the amount of coherence in the initial state is shown in \fig{fig:workentropycoherence}, where coherence of the state is calculated by \eq{equ:relativecoherence}. Here cyan dots represent states with two non-zero diagonal terms and random coherence, and green dots represent coherent thermal states $\hr_{S}$ with random coherence. Again we see that coherence does not help produce work in \fig{fig:coherenceandwork}, while it monotonically increases the irreversible entropy production in \fig{fig:coherenceandentropy}. 


We are now interested in the effects of model parameters on the work production of the process. Given \eq{equ:HeisenbergHam}, we assume without loss of generality that $\epsilon_1>\epsilon_2> 0$ and $\alpha,\beta,\gamma$ all positive. Their values are all set to be $1$ unless specified otherwise. The instantaneous eigenvalues of the system Hamiltonian are then $\{E_s^1,E_s^2,E_s^3,E_s^4\}$, given in \eq{equ:tdeigenvalues}, with $E_0^j<E_0^{j+1}~(j=1,..3)$. Due to the symmetry of the model, we look at the effects of $\alpha, \gamma$ on the work production of the process.

Instead of giving the full picture as in \fig{fig:workandentropyallstates}, here we focus on two states. One that gives the maximum work extraction, and the other state that produces zero irreversible entropy. The maximum work extraction is given the following expression, \mauro{sentence unclear}
\begin{equation}
\Delta W_{max} = -\max_i (E_0^i-E_\tau^i).
\end{equation}
Studying how the eigenvalues $\{E_s^1,E_s^2,E_s^3,E_s^4\}$ change in different regimes, we find that the state that provides maximum work extraction is always $\hr_{max}=diag(0,0,1,0)$. Namely, the maximum work output is alway $E_\tau^3-E_0^3$. On the other hand, $\{E_\tau^1,E_\tau^2,E_\tau^3,E_\tau^4\}$ determine the final hypothetical thermal state $\hr_{th}=diag(\rho_{th}^1,\rho_{th}^2,\rho_{th}^3,\rho_{th}^4)$ with $\rho_{th}^i=\exp(-\beta E_\tau^i)/Z(\beta)$. To produce zero irreversible entropy, the initial state $\hr_{in}$ has to have the same populations as $\hr_{th}$ in the beginning basis, with some populations being permuted due to energy crossings.

Effects of changing parameters $\alpha$ and $\gamma$ can be seen in \fig{fig:maximumworkproduction}. Suppose the initial state is $\hr_{max}$ 
(shown in Fig.~(\ref{fig:workentropyalphagamma})), increasing either of parameters monotonically increases the maximum work extraction, meanwhile reduces the irreversible entropy production. However, irreversible entropy cannot be reduced to zero for this initial state, but reaching to the limit $-\ln\left(\frac{1}{1+e^{-2}}\right)$, shown by the dashed line. On the other hand, suppose the initial state is prepared so as to produce zero irreversible entropy (shown in Fig.~(\ref{fig:workalphagamma})), increasing either of parameters also increases the maximum work extraction. One can see there exists a jump in work production in the process of increasing $\gamma$. This happens because energy levels of the final Hamiltonian cross each other at this value. After this point, no jump is observed when $\gamma$ keeps growing. Such energy crossing also occurs when increasing the value of $\gamma$. Interestingly, there is no jump observed when the crossing happens in the process of increasing $\alpha$. \mauro{unclear}






\section{Conclusions}\label{sec:conclusion}

We have studied the relation between work extracted and irreversible entropy production following a process assisted by a TQD protocol. We have focused on the role played by quantum coherence in the initial state of the work medium -- which has been taken to be either a single- or a two-qubit interacting system  -- and identified the states that extremize the amount of work extracted at a given degree of entropy production. Our study provides a systematic assessment of the role of initial state preparation in the performance of an out-of-equilibrium process enhanced by quantum control protocols, which is a situation of strong experimental relevance. We believe that our results will be useful to inform the next generation of experiments addressing the thermodynamics of quantum working media.

\section{Acknowledgements}
The authors would like to thank Obinna Abah and Gabriel T. Landi for stimulating discussions. They acknowledge financial support from H2020 through Collaborative Project TEQ (Grant Agreement No.  766900), the DfE-SFI Investigator Programme (Grant No. 15/IA/2864), the Leverhulme Trust Research Project Grant UltraQute (grant nr.~RGP-2018-266), COST Action CA15220, and the Royal Society Wolfson Research Fellowship scheme (RSWF\textbackslash R3\textbackslash183013).

\bibliographystyle{apsrev4-1}
\bibliography{refers}

\begin{thebibliography}{46}%
\makeatletter
\providecommand \@ifxundefined [1]{%
 \@ifx{#1\undefined}
}%
\providecommand \@ifnum [1]{%
 \ifnum #1\expandafter \@firstoftwo
 \else \expandafter \@secondoftwo
 \fi
}%
\providecommand \@ifx [1]{%
 \ifx #1\expandafter \@firstoftwo
 \else \expandafter \@secondoftwo
 \fi
}%
\providecommand \natexlab [1]{#1}%
\providecommand \enquote  [1]{``#1''}%
\providecommand \bibnamefont  [1]{#1}%
\providecommand \bibfnamefont [1]{#1}%
\providecommand \citenamefont [1]{#1}%
\providecommand \href@noop [0]{\@secondoftwo}%
\providecommand \href [0]{\begingroup \@sanitize@url \@href}%
\providecommand \@href[1]{\@@startlink{#1}\@@href}%
\providecommand \@@href[1]{\endgroup#1\@@endlink}%
\providecommand \@sanitize@url [0]{\catcode `\\12\catcode `\$12\catcode
  `\&12\catcode `\#12\catcode `\^12\catcode `\_12\catcode `\%12\relax}%
\providecommand \@@startlink[1]{}%
\providecommand \@@endlink[0]{}%
\providecommand \url  [0]{\begingroup\@sanitize@url \@url }%
\providecommand \@url [1]{\endgroup\@href {#1}{\urlprefix }}%
\providecommand \urlprefix  [0]{URL }%
\providecommand \Eprint [0]{\href }%
\providecommand \doibase [0]{https://doi.org/}%
\providecommand \selectlanguage [0]{\@gobble}%
\providecommand \bibinfo  [0]{\@secondoftwo}%
\providecommand \bibfield  [0]{\@secondoftwo}%
\providecommand \translation [1]{[#1]}%
\providecommand \BibitemOpen [0]{}%
\providecommand \bibitemStop [0]{}%
\providecommand \bibitemNoStop [0]{.\EOS\space}%
\providecommand \EOS [0]{\spacefactor3000\relax}%
\providecommand \BibitemShut  [1]{\csname bibitem#1\endcsname}%
\let\auto@bib@innerbib\@empty
\bibitem [{\citenamefont {Kosloff}(2013)}]{Kosloff_2013}%
  \BibitemOpen
  \bibfield  {author} {\bibinfo {author} {\bibfnamefont {R.}~\bibnamefont
  {Kosloff}},\ }\href {https://doi.org/10.3390/e15062100} {\bibfield  {journal}
  {\bibinfo  {journal} {Entropy}\ }\textbf {\bibinfo {volume} {15}},\ \bibinfo
  {pages} {2100} (\bibinfo {year} {2013})}\BibitemShut {NoStop}%
\bibitem [{\citenamefont {Brand{\~a}o}\ \emph {et~al.}(2015)\citenamefont
  {Brand{\~a}o}, \citenamefont {Horodecki}, \citenamefont {Ng}, \citenamefont
  {Oppenheim},\ and\ \citenamefont {Wehner}}]{Brandao_2015}%
  \BibitemOpen
  \bibfield  {author} {\bibinfo {author} {\bibfnamefont {F.}~\bibnamefont
  {Brand{\~a}o}}, \bibinfo {author} {\bibfnamefont {M.}~\bibnamefont
  {Horodecki}}, \bibinfo {author} {\bibfnamefont {N.}~\bibnamefont {Ng}},
  \bibinfo {author} {\bibfnamefont {J.}~\bibnamefont {Oppenheim}},\ and\
  \bibinfo {author} {\bibfnamefont {S.}~\bibnamefont {Wehner}},\ }\href
  {https://doi.org/10.1073/pnas.1411728112} {\bibfield  {journal} {\bibinfo
  {journal} {Proceedings of the National Academy of Sciences}\ }\textbf
  {\bibinfo {volume} {112}},\ \bibinfo {pages} {3275} (\bibinfo {year}
  {2015})}\BibitemShut {NoStop}%
\bibitem [{\citenamefont {{Vinjanampathy}}\ and\ \citenamefont
  {{Anders}}(2016)}]{Vinjanampathy_2016}%
  \BibitemOpen
  \bibfield  {author} {\bibinfo {author} {\bibfnamefont {S.}~\bibnamefont
  {{Vinjanampathy}}}\ and\ \bibinfo {author} {\bibfnamefont {J.}~\bibnamefont
  {{Anders}}},\ }\href {https://doi.org/10.1080/00107514.2016.1201896}
  {\bibfield  {journal} {\bibinfo  {journal} {Contemporary Physics}\ }\textbf
  {\bibinfo {volume} {57}},\ \bibinfo {pages} {545} (\bibinfo {year}
  {2016})}\BibitemShut {NoStop}%
\bibitem [{\citenamefont {Horodecki}\ and\ \citenamefont
  {Oppenheim}(2013)}]{Horodecki_2013}%
  \BibitemOpen
  \bibfield  {author} {\bibinfo {author} {\bibfnamefont {M.}~\bibnamefont
  {Horodecki}}\ and\ \bibinfo {author} {\bibfnamefont {J.}~\bibnamefont
  {Oppenheim}},\ }\href {https://doi.org/10.1038/ncomms3059} {\bibfield
  {journal} {\bibinfo  {journal} {Nature Communications}\ }\textbf {\bibinfo
  {volume} {4}},\ \bibinfo {pages} {2059} (\bibinfo {year} {2013})}\BibitemShut
  {NoStop}%
\bibitem [{\citenamefont {Rio}\ \emph {et~al.}(2011)\citenamefont {Rio},
  \citenamefont {{\AA}berg}, \citenamefont {Renner}, \citenamefont {Dahlsten},\
  and\ \citenamefont {Vedral}}]{Rio_2011}%
  \BibitemOpen
  \bibfield  {author} {\bibinfo {author} {\bibfnamefont {L.~d.}\ \bibnamefont
  {Rio}}, \bibinfo {author} {\bibfnamefont {J.}~\bibnamefont {{\AA}berg}},
  \bibinfo {author} {\bibfnamefont {R.}~\bibnamefont {Renner}}, \bibinfo
  {author} {\bibfnamefont {O.}~\bibnamefont {Dahlsten}},\ and\ \bibinfo
  {author} {\bibfnamefont {V.}~\bibnamefont {Vedral}},\ }\href
  {https://doi.org/10.1038/nature10123} {\bibfield  {journal} {\bibinfo
  {journal} {Nature}\ }\textbf {\bibinfo {volume} {474}},\ \bibinfo {pages}
  {61} (\bibinfo {year} {2011})}\BibitemShut {NoStop}%
\bibitem [{\citenamefont {Hsiang}\ \emph {et~al.}(2018)\citenamefont {Hsiang},
  \citenamefont {Chou}, \citenamefont {Suba\ifmmode \mbox{\c{s}}\else
  \c{s}\fi{}\ifmmode \imath \else~\i \fi{}},\ and\ \citenamefont
  {Hu}}]{Hsiang_2018}%
  \BibitemOpen
  \bibfield  {author} {\bibinfo {author} {\bibfnamefont {J.-T.}\ \bibnamefont
  {Hsiang}}, \bibinfo {author} {\bibfnamefont {C.~H.}\ \bibnamefont {Chou}},
  \bibinfo {author} {\bibfnamefont {Y.}~\bibnamefont {Suba\ifmmode
  \mbox{\c{s}}\else \c{s}\fi{}\ifmmode \imath \else~\i \fi{}}},\ and\ \bibinfo
  {author} {\bibfnamefont {B.~L.}\ \bibnamefont {Hu}},\ }\href
  {https://doi.org/10.1103/PhysRevE.97.012135} {\bibfield  {journal} {\bibinfo
  {journal} {Phys. Rev. E}\ }\textbf {\bibinfo {volume} {97}},\ \bibinfo
  {pages} {012135} (\bibinfo {year} {2018})}\BibitemShut {NoStop}%
\bibitem [{\citenamefont {Santos}\ \emph {et~al.}(2019)\citenamefont {Santos},
  \citenamefont {C{\'e}leri}, \citenamefont {Landi},\ and\ \citenamefont
  {Paternostro}}]{Santos_2019}%
  \BibitemOpen
  \bibfield  {author} {\bibinfo {author} {\bibfnamefont {J.~P.}\ \bibnamefont
  {Santos}}, \bibinfo {author} {\bibfnamefont {L.~C.}\ \bibnamefont
  {C{\'e}leri}}, \bibinfo {author} {\bibfnamefont {G.~T.}\ \bibnamefont
  {Landi}},\ and\ \bibinfo {author} {\bibfnamefont {M.}~\bibnamefont
  {Paternostro}},\ }\href {https://doi.org/10.1038/s41534-019-0138-y}
  {\bibfield  {journal} {\bibinfo  {journal} {npj Quantum Information}\
  }\textbf {\bibinfo {volume} {5}},\ \bibinfo {pages} {23} (\bibinfo {year}
  {2019})}\BibitemShut {NoStop}%
\bibitem [{\citenamefont {Kim}\ \emph {et~al.}(2011)\citenamefont {Kim},
  \citenamefont {Sagawa}, \citenamefont {De~Liberato},\ and\ \citenamefont
  {Ueda}}]{Kim_2011}%
  \BibitemOpen
  \bibfield  {author} {\bibinfo {author} {\bibfnamefont {S.~W.}\ \bibnamefont
  {Kim}}, \bibinfo {author} {\bibfnamefont {T.}~\bibnamefont {Sagawa}},
  \bibinfo {author} {\bibfnamefont {S.}~\bibnamefont {De~Liberato}},\ and\
  \bibinfo {author} {\bibfnamefont {M.}~\bibnamefont {Ueda}},\ }\href
  {https://doi.org/10.1103/PhysRevLett.106.070401} {\bibfield  {journal}
  {\bibinfo  {journal} {Phys. Rev. Lett.}\ }\textbf {\bibinfo {volume} {106}},\
  \bibinfo {pages} {070401} (\bibinfo {year} {2011})}\BibitemShut {NoStop}%
\bibitem [{\citenamefont {Ro{\ss}nagel}\ \emph {et~al.}(2016)\citenamefont
  {Ro{\ss}nagel}, \citenamefont {Dawkins}, \citenamefont {Tolazzi},
  \citenamefont {Abah}, \citenamefont {Lutz}, \citenamefont {Schmidt-Kaler},\
  and\ \citenamefont {Singer}}]{Rossnagel_2016}%
  \BibitemOpen
  \bibfield  {author} {\bibinfo {author} {\bibfnamefont {J.}~\bibnamefont
  {Ro{\ss}nagel}}, \bibinfo {author} {\bibfnamefont {S.~T.}\ \bibnamefont
  {Dawkins}}, \bibinfo {author} {\bibfnamefont {K.~N.}\ \bibnamefont
  {Tolazzi}}, \bibinfo {author} {\bibfnamefont {O.}~\bibnamefont {Abah}},
  \bibinfo {author} {\bibfnamefont {E.}~\bibnamefont {Lutz}}, \bibinfo {author}
  {\bibfnamefont {F.}~\bibnamefont {Schmidt-Kaler}},\ and\ \bibinfo {author}
  {\bibfnamefont {K.}~\bibnamefont {Singer}},\ }\href
  {https://doi.org/10.1126/science.aad6320} {\bibfield  {journal} {\bibinfo
  {journal} {Science}\ }\textbf {\bibinfo {volume} {352}},\ \bibinfo {pages}
  {325} (\bibinfo {year} {2016})}\BibitemShut {NoStop}%
\bibitem [{\citenamefont {Abah}\ and\ \citenamefont {Lutz}(2017)}]{Abah_2017}%
  \BibitemOpen
  \bibfield  {author} {\bibinfo {author} {\bibfnamefont {O.}~\bibnamefont
  {Abah}}\ and\ \bibinfo {author} {\bibfnamefont {E.}~\bibnamefont {Lutz}},\
  }\href {https://doi.org/10.1209/0295-5075/118/40005} {\bibfield  {journal}
  {\bibinfo  {journal} {{EPL} (Europhysics Letters)}\ }\textbf {\bibinfo
  {volume} {118}},\ \bibinfo {pages} {40005} (\bibinfo {year}
  {2017})}\BibitemShut {NoStop}%
\bibitem [{\citenamefont {Goold}\ \emph {et~al.}(2016)\citenamefont {Goold},
  \citenamefont {Huber}, \citenamefont {Riera}, \citenamefont {del Rio},\ and\
  \citenamefont {Skrzypczyk}}]{Goold_2016}%
  \BibitemOpen
  \bibfield  {author} {\bibinfo {author} {\bibfnamefont {J.}~\bibnamefont
  {Goold}}, \bibinfo {author} {\bibfnamefont {M.}~\bibnamefont {Huber}},
  \bibinfo {author} {\bibfnamefont {A.}~\bibnamefont {Riera}}, \bibinfo
  {author} {\bibfnamefont {L.}~\bibnamefont {del Rio}},\ and\ \bibinfo {author}
  {\bibfnamefont {P.}~\bibnamefont {Skrzypczyk}},\ }\href
  {https://doi.org/10.1088/1751-8113/49/14/143001} {\bibfield  {journal}
  {\bibinfo  {journal} {Journal of Physics A: Mathematical and Theoretical}\
  }\textbf {\bibinfo {volume} {49}},\ \bibinfo {pages} {143001} (\bibinfo
  {year} {2016})}\BibitemShut {NoStop}%
\bibitem [{\citenamefont {{Deffner}}\ and\ \citenamefont
  {{Campbell}}(2019)}]{Deffner_2019}%
  \BibitemOpen
  \bibfield  {author} {\bibinfo {author} {\bibfnamefont {S.}~\bibnamefont
  {{Deffner}}}\ and\ \bibinfo {author} {\bibfnamefont {S.}~\bibnamefont
  {{Campbell}}},\ }\href@noop {} {\bibfield  {journal} {\bibinfo  {journal}
  {arXiv e-prints}\ ,\ \bibinfo {eid} {arXiv:1907.01596}} (\bibinfo {year}
  {2019})},\ \Eprint {https://arxiv.org/abs/1907.01596} {arXiv:1907.01596
  [quant-ph]} \BibitemShut {NoStop}%
\bibitem [{\citenamefont {Liphardt}\ \emph {et~al.}(2002)\citenamefont
  {Liphardt}, \citenamefont {Dumont}, \citenamefont {Smith}, \citenamefont
  {Tinoco},\ and\ \citenamefont {Bustamante}}]{Liphardt_2002}%
  \BibitemOpen
  \bibfield  {author} {\bibinfo {author} {\bibfnamefont {J.}~\bibnamefont
  {Liphardt}}, \bibinfo {author} {\bibfnamefont {S.}~\bibnamefont {Dumont}},
  \bibinfo {author} {\bibfnamefont {S.~B.}\ \bibnamefont {Smith}}, \bibinfo
  {author} {\bibfnamefont {I.}~\bibnamefont {Tinoco}},\ and\ \bibinfo {author}
  {\bibfnamefont {C.}~\bibnamefont {Bustamante}},\ }\href
  {https://doi.org/10.1126/science.1071152} {\bibfield  {journal} {\bibinfo
  {journal} {Science}\ }\textbf {\bibinfo {volume} {296}},\ \bibinfo {pages}
  {1832} (\bibinfo {year} {2002})}\BibitemShut {NoStop}%
\bibitem [{\citenamefont {An}\ \emph {et~al.}(2015)\citenamefont {An},
  \citenamefont {Zhang}, \citenamefont {Um}, \citenamefont {Lv}, \citenamefont
  {Lu}, \citenamefont {Zhang}, \citenamefont {Yin}, \citenamefont {Quan},\ and\
  \citenamefont {Kim}}]{An_2015}%
  \BibitemOpen
  \bibfield  {author} {\bibinfo {author} {\bibfnamefont {S.}~\bibnamefont
  {An}}, \bibinfo {author} {\bibfnamefont {J.-N.}\ \bibnamefont {Zhang}},
  \bibinfo {author} {\bibfnamefont {M.}~\bibnamefont {Um}}, \bibinfo {author}
  {\bibfnamefont {D.}~\bibnamefont {Lv}}, \bibinfo {author} {\bibfnamefont
  {Y.}~\bibnamefont {Lu}}, \bibinfo {author} {\bibfnamefont {J.}~\bibnamefont
  {Zhang}}, \bibinfo {author} {\bibfnamefont {Z.-Q.}\ \bibnamefont {Yin}},
  \bibinfo {author} {\bibfnamefont {H.~T.}\ \bibnamefont {Quan}},\ and\
  \bibinfo {author} {\bibfnamefont {K.}~\bibnamefont {Kim}},\ }\href
  {https://doi.org/10.1038/nphys3197} {\bibfield  {journal} {\bibinfo
  {journal} {Nature Physics}\ }\textbf {\bibinfo {volume} {11}},\ \bibinfo
  {pages} {193} (\bibinfo {year} {2015})}\BibitemShut {NoStop}%
\bibitem [{\citenamefont {de~Ara{\'{u}}jo}\ \emph {et~al.}(2018)\citenamefont
  {de~Ara{\'{u}}jo}, \citenamefont {Häffner}, \citenamefont {Bernardi},
  \citenamefont {Tasca}, \citenamefont {Lavery}, \citenamefont {Padgett},
  \citenamefont {Kanaan}, \citenamefont {C{\'{e}}leri},\ and\ \citenamefont
  {Ribeiro}}]{Ara_jo_2018}%
  \BibitemOpen
  \bibfield  {author} {\bibinfo {author} {\bibfnamefont {R.~M.}\ \bibnamefont
  {de~Ara{\'{u}}jo}}, \bibinfo {author} {\bibfnamefont {T.}~\bibnamefont
  {Häffner}}, \bibinfo {author} {\bibfnamefont {R.}~\bibnamefont {Bernardi}},
  \bibinfo {author} {\bibfnamefont {D.~S.}\ \bibnamefont {Tasca}}, \bibinfo
  {author} {\bibfnamefont {M.~P.~J.}\ \bibnamefont {Lavery}}, \bibinfo {author}
  {\bibfnamefont {M.~J.}\ \bibnamefont {Padgett}}, \bibinfo {author}
  {\bibfnamefont {A.}~\bibnamefont {Kanaan}}, \bibinfo {author} {\bibfnamefont
  {L.~C.}\ \bibnamefont {C{\'{e}}leri}},\ and\ \bibinfo {author} {\bibfnamefont
  {P.~H.~S.}\ \bibnamefont {Ribeiro}},\ }\href
  {https://doi.org/10.1088/2399-6528/aab178} {\bibfield  {journal} {\bibinfo
  {journal} {Journal of Physics Communications}\ }\textbf {\bibinfo {volume}
  {2}},\ \bibinfo {pages} {035012} (\bibinfo {year} {2018})}\BibitemShut
  {NoStop}%
\bibitem [{\citenamefont {Klatzow}\ \emph {et~al.}(2019)\citenamefont
  {Klatzow}, \citenamefont {Becker}, \citenamefont {Ledingham}, \citenamefont
  {Weinzetl}, \citenamefont {Kaczmarek}, \citenamefont {Saunders},
  \citenamefont {Nunn}, \citenamefont {Walmsley}, \citenamefont {Uzdin},\ and\
  \citenamefont {Poem}}]{Klatzow_2019}%
  \BibitemOpen
  \bibfield  {author} {\bibinfo {author} {\bibfnamefont {J.}~\bibnamefont
  {Klatzow}}, \bibinfo {author} {\bibfnamefont {J.~N.}\ \bibnamefont {Becker}},
  \bibinfo {author} {\bibfnamefont {P.~M.}\ \bibnamefont {Ledingham}}, \bibinfo
  {author} {\bibfnamefont {C.}~\bibnamefont {Weinzetl}}, \bibinfo {author}
  {\bibfnamefont {K.~T.}\ \bibnamefont {Kaczmarek}}, \bibinfo {author}
  {\bibfnamefont {D.~J.}\ \bibnamefont {Saunders}}, \bibinfo {author}
  {\bibfnamefont {J.}~\bibnamefont {Nunn}}, \bibinfo {author} {\bibfnamefont
  {I.~A.}\ \bibnamefont {Walmsley}}, \bibinfo {author} {\bibfnamefont
  {R.}~\bibnamefont {Uzdin}},\ and\ \bibinfo {author} {\bibfnamefont
  {E.}~\bibnamefont {Poem}},\ }\href
  {https://doi.org/10.1103/PhysRevLett.122.110601} {\bibfield  {journal}
  {\bibinfo  {journal} {Phys. Rev. Lett.}\ }\textbf {\bibinfo {volume} {122}},\
  \bibinfo {pages} {110601} (\bibinfo {year} {2019})}\BibitemShut {NoStop}%
\bibitem [{\citenamefont {Von~Lindenfels}\ \emph {et~al.}(2019)\citenamefont
  {Von~Lindenfels}, \citenamefont {Gr\"ab}, \citenamefont {Schmiegelow},
  \citenamefont {Kaushal}, \citenamefont {Schulz}, \citenamefont {Mitchison},
  \citenamefont {Goold}, \citenamefont {Schmidt-Kaler},\ and\ \citenamefont
  {Poschinger}}]{von_Lindenfels_2019}%
  \BibitemOpen
  \bibfield  {author} {\bibinfo {author} {\bibfnamefont {D.}~\bibnamefont
  {Von~Lindenfels}}, \bibinfo {author} {\bibfnamefont {O.}~\bibnamefont
  {Gr\"ab}}, \bibinfo {author} {\bibfnamefont {C.~T.}\ \bibnamefont
  {Schmiegelow}}, \bibinfo {author} {\bibfnamefont {V.}~\bibnamefont
  {Kaushal}}, \bibinfo {author} {\bibfnamefont {J.}~\bibnamefont {Schulz}},
  \bibinfo {author} {\bibfnamefont {M.~T.}\ \bibnamefont {Mitchison}}, \bibinfo
  {author} {\bibfnamefont {J.}~\bibnamefont {Goold}}, \bibinfo {author}
  {\bibfnamefont {F.}~\bibnamefont {Schmidt-Kaler}},\ and\ \bibinfo {author}
  {\bibfnamefont {U.~G.}\ \bibnamefont {Poschinger}},\ }\href
  {https://doi.org/10.1103/PhysRevLett.123.080602} {\bibfield  {journal}
  {\bibinfo  {journal} {Phys. Rev. Lett.}\ }\textbf {\bibinfo {volume} {123}},\
  \bibinfo {pages} {080602} (\bibinfo {year} {2019})}\BibitemShut {NoStop}%
\bibitem [{\citenamefont {Peterson}\ \emph {et~al.}(2019)\citenamefont
  {Peterson}, \citenamefont {Batalh\~ao}, \citenamefont {Herrera},
  \citenamefont {Souza}, \citenamefont {Sarthour}, \citenamefont {Oliveira},\
  and\ \citenamefont {Serra}}]{Peterson_2019}%
  \BibitemOpen
  \bibfield  {author} {\bibinfo {author} {\bibfnamefont {J.~P.~S.}\
  \bibnamefont {Peterson}}, \bibinfo {author} {\bibfnamefont {T.~B.}\
  \bibnamefont {Batalh\~ao}}, \bibinfo {author} {\bibfnamefont
  {M.}~\bibnamefont {Herrera}}, \bibinfo {author} {\bibfnamefont {A.~M.}\
  \bibnamefont {Souza}}, \bibinfo {author} {\bibfnamefont {R.~S.}\ \bibnamefont
  {Sarthour}}, \bibinfo {author} {\bibfnamefont {I.~S.}\ \bibnamefont
  {Oliveira}},\ and\ \bibinfo {author} {\bibfnamefont {R.~M.}\ \bibnamefont
  {Serra}},\ }\href {https://doi.org/10.1103/PhysRevLett.123.240601} {\bibfield
   {journal} {\bibinfo  {journal} {Phys. Rev. Lett.}\ }\textbf {\bibinfo
  {volume} {123}},\ \bibinfo {pages} {240601} (\bibinfo {year}
  {2019})}\BibitemShut {NoStop}%
\bibitem [{\citenamefont {Feldmann}\ and\ \citenamefont
  {Kosloff}(2003)}]{Feldmann_2003}%
  \BibitemOpen
  \bibfield  {author} {\bibinfo {author} {\bibfnamefont {T.}~\bibnamefont
  {Feldmann}}\ and\ \bibinfo {author} {\bibfnamefont {R.}~\bibnamefont
  {Kosloff}},\ }\href {https://doi.org/10.1103/PhysRevE.68.016101} {\bibfield
  {journal} {\bibinfo  {journal} {Phys. Rev. E}\ }\textbf {\bibinfo {volume}
  {68}},\ \bibinfo {pages} {016101} (\bibinfo {year} {2003})}\BibitemShut
  {NoStop}%
\bibitem [{\citenamefont {Quan}\ \emph {et~al.}(2007)\citenamefont {Quan},
  \citenamefont {Liu}, \citenamefont {Sun},\ and\ \citenamefont
  {Nori}}]{Quan_2007}%
  \BibitemOpen
  \bibfield  {author} {\bibinfo {author} {\bibfnamefont {H.~T.}\ \bibnamefont
  {Quan}}, \bibinfo {author} {\bibfnamefont {Y.-x.}\ \bibnamefont {Liu}},
  \bibinfo {author} {\bibfnamefont {C.~P.}\ \bibnamefont {Sun}},\ and\ \bibinfo
  {author} {\bibfnamefont {F.}~\bibnamefont {Nori}},\ }\href
  {https://doi.org/10.1103/PhysRevE.76.031105} {\bibfield  {journal} {\bibinfo
  {journal} {Phys. Rev. E}\ }\textbf {\bibinfo {volume} {76}},\ \bibinfo
  {pages} {031105} (\bibinfo {year} {2007})}\BibitemShut {NoStop}%
\bibitem [{\citenamefont {Quan}(2009)}]{Quan_2009}%
  \BibitemOpen
  \bibfield  {author} {\bibinfo {author} {\bibfnamefont {H.~T.}\ \bibnamefont
  {Quan}},\ }\href {https://doi.org/10.1103/PhysRevE.79.041129} {\bibfield
  {journal} {\bibinfo  {journal} {Phys. Rev. E}\ }\textbf {\bibinfo {volume}
  {79}},\ \bibinfo {pages} {041129} (\bibinfo {year} {2009})}\BibitemShut
  {NoStop}%
\bibitem [{\citenamefont {Barontini}\ and\ \citenamefont
  {Paternostro}(2019)}]{Barontini_2019}%
  \BibitemOpen
  \bibfield  {author} {\bibinfo {author} {\bibfnamefont {G.}~\bibnamefont
  {Barontini}}\ and\ \bibinfo {author} {\bibfnamefont {M.}~\bibnamefont
  {Paternostro}},\ }\href {https://doi.org/10.1088/1367-2630/ab2684} {\bibfield
   {journal} {\bibinfo  {journal} {New Journal of Physics}\ }\textbf {\bibinfo
  {volume} {21}},\ \bibinfo {pages} {063019} (\bibinfo {year}
  {2019})}\BibitemShut {NoStop}%
\bibitem [{\citenamefont {Scully}(2002)}]{Scully_2002}%
  \BibitemOpen
  \bibfield  {author} {\bibinfo {author} {\bibfnamefont {M.~O.}\ \bibnamefont
  {Scully}},\ }\href {https://doi.org/10.1103/PhysRevLett.88.050602} {\bibfield
   {journal} {\bibinfo  {journal} {Phys. Rev. Lett.}\ }\textbf {\bibinfo
  {volume} {88}},\ \bibinfo {pages} {050602} (\bibinfo {year}
  {2002})}\BibitemShut {NoStop}%
\bibitem [{\citenamefont {Dillenschneider}\ and\ \citenamefont
  {Lutz}(2009)}]{Dillenschneider_2009}%
  \BibitemOpen
  \bibfield  {author} {\bibinfo {author} {\bibfnamefont {R.}~\bibnamefont
  {Dillenschneider}}\ and\ \bibinfo {author} {\bibfnamefont {E.}~\bibnamefont
  {Lutz}},\ }\href {https://doi.org/10.1209/0295-5075/88/50003} {\bibfield
  {journal} {\bibinfo  {journal} {{EPL} (Europhysics Letters)}\ }\textbf
  {\bibinfo {volume} {88}},\ \bibinfo {pages} {50003} (\bibinfo {year}
  {2009})}\BibitemShut {NoStop}%
\bibitem [{\citenamefont {Scully}\ \emph {et~al.}(2011)\citenamefont {Scully},
  \citenamefont {Chapin}, \citenamefont {Dorfman}, \citenamefont {Kim},\ and\
  \citenamefont {Svidzinsky}}]{Scully_2011}%
  \BibitemOpen
  \bibfield  {author} {\bibinfo {author} {\bibfnamefont {M.~O.}\ \bibnamefont
  {Scully}}, \bibinfo {author} {\bibfnamefont {K.~R.}\ \bibnamefont {Chapin}},
  \bibinfo {author} {\bibfnamefont {K.~E.}\ \bibnamefont {Dorfman}}, \bibinfo
  {author} {\bibfnamefont {M.~B.}\ \bibnamefont {Kim}},\ and\ \bibinfo {author}
  {\bibfnamefont {A.}~\bibnamefont {Svidzinsky}},\ }\href
  {https://doi.org/10.1073/pnas.1110234108} {\bibfield  {journal} {\bibinfo
  {journal} {Proceedings of the National Academy of Sciences}\ }\textbf
  {\bibinfo {volume} {108}},\ \bibinfo {pages} {15097} (\bibinfo {year}
  {2011})}\BibitemShut {NoStop}%
\bibitem [{\citenamefont {Berry}(2009)}]{Berry_2009}%
  \BibitemOpen
  \bibfield  {author} {\bibinfo {author} {\bibfnamefont {M.}~\bibnamefont
  {Berry}},\ }\href {https://doi.org/10.1088/1751-8113/42/36/365303} {\bibfield
   {journal} {\bibinfo  {journal} {Journal of Physics A: Mathematical and
  Theoretical}\ }\textbf {\bibinfo {volume} {42}},\ \bibinfo {pages} {365303}
  (\bibinfo {year} {2009})}\BibitemShut {NoStop}%
\bibitem [{\citenamefont {Demirplak}\ and\ \citenamefont
  {Rice}(2003)}]{Demirplak_2003}%
  \BibitemOpen
  \bibfield  {author} {\bibinfo {author} {\bibfnamefont {M.}~\bibnamefont
  {Demirplak}}\ and\ \bibinfo {author} {\bibfnamefont {S.~A.}\ \bibnamefont
  {Rice}},\ }\href {https://doi.org/10.1021/jp030708a} {\bibfield  {journal}
  {\bibinfo  {journal} {The Journal of Physical Chemistry A}\ }\textbf
  {\bibinfo {volume} {107}},\ \bibinfo {pages} {9937} (\bibinfo {year}
  {2003})}\BibitemShut {NoStop}%
\bibitem [{\citenamefont {Demirplak}\ and\ \citenamefont
  {Rice}(2005)}]{Demirplak_2005}%
  \BibitemOpen
  \bibfield  {author} {\bibinfo {author} {\bibfnamefont {M.}~\bibnamefont
  {Demirplak}}\ and\ \bibinfo {author} {\bibfnamefont {S.~A.}\ \bibnamefont
  {Rice}},\ }\href {https://doi.org/10.1021/jp040647w} {\bibfield  {journal}
  {\bibinfo  {journal} {The Journal of Physical Chemistry B}\ }\textbf
  {\bibinfo {volume} {109}},\ \bibinfo {pages} {6838} (\bibinfo {year}
  {2005})},\ \bibinfo {note} {pMID: 16851769}\BibitemShut {NoStop}%
\bibitem [{\citenamefont {Demirplak}\ and\ \citenamefont
  {Rice}(2008)}]{Demirplak_2008}%
  \BibitemOpen
  \bibfield  {author} {\bibinfo {author} {\bibfnamefont {M.}~\bibnamefont
  {Demirplak}}\ and\ \bibinfo {author} {\bibfnamefont {S.~A.}\ \bibnamefont
  {Rice}},\ }\href {https://doi.org/10.1063/1.2992152} {\bibfield  {journal}
  {\bibinfo  {journal} {The Journal of Chemical Physics}\ }\textbf {\bibinfo
  {volume} {129}},\ \bibinfo {pages} {154111} (\bibinfo {year}
  {2008})}\BibitemShut {NoStop}%
\bibitem [{\citenamefont {Gu\'ery-Odelin}\ \emph {et~al.}(2019)\citenamefont
  {Gu\'ery-Odelin}, \citenamefont {Ruschhaupt}, \citenamefont {Kiely},
  \citenamefont {Torrontegui}, \citenamefont {Mart\'{\i}nez-Garaot},\ and\
  \citenamefont {Muga}}]{GueryOdelin_2019}%
  \BibitemOpen
  \bibfield  {author} {\bibinfo {author} {\bibfnamefont {D.}~\bibnamefont
  {Gu\'ery-Odelin}}, \bibinfo {author} {\bibfnamefont {A.}~\bibnamefont
  {Ruschhaupt}}, \bibinfo {author} {\bibfnamefont {A.}~\bibnamefont {Kiely}},
  \bibinfo {author} {\bibfnamefont {E.}~\bibnamefont {Torrontegui}}, \bibinfo
  {author} {\bibfnamefont {S.}~\bibnamefont {Mart\'{\i}nez-Garaot}},\ and\
  \bibinfo {author} {\bibfnamefont {J.~G.}\ \bibnamefont {Muga}},\ }\href
  {https://doi.org/10.1103/RevModPhys.91.045001} {\bibfield  {journal}
  {\bibinfo  {journal} {Rev. Mod. Phys.}\ }\textbf {\bibinfo {volume} {91}},\
  \bibinfo {pages} {045001} (\bibinfo {year} {2019})}\BibitemShut {NoStop}%
\bibitem [{\citenamefont {del Campo}(2013)}]{del_Campo_2013}%
  \BibitemOpen
  \bibfield  {author} {\bibinfo {author} {\bibfnamefont {A.}~\bibnamefont {del
  Campo}},\ }\href {https://doi.org/10.1103/PhysRevLett.111.100502} {\bibfield
  {journal} {\bibinfo  {journal} {Phys. Rev. Lett.}\ }\textbf {\bibinfo
  {volume} {111}},\ \bibinfo {pages} {100502} (\bibinfo {year}
  {2013})}\BibitemShut {NoStop}%
\bibitem [{\citenamefont {del Campo}\ \emph {et~al.}(2014)\citenamefont {del
  Campo}, \citenamefont {Goold},\ and\ \citenamefont
  {Paternostro}}]{Campo_2014}%
  \BibitemOpen
  \bibfield  {author} {\bibinfo {author} {\bibfnamefont {A.}~\bibnamefont {del
  Campo}}, \bibinfo {author} {\bibfnamefont {J.}~\bibnamefont {Goold}},\ and\
  \bibinfo {author} {\bibfnamefont {M.}~\bibnamefont {Paternostro}},\ }\href
  {https://doi.org/10.1038/srep06208} {\bibfield  {journal} {\bibinfo
  {journal} {Scientific Reports}\ }\textbf {\bibinfo {volume} {4}},\ \bibinfo
  {pages} {6208} (\bibinfo {year} {2014})}\BibitemShut {NoStop}%
\bibitem [{\citenamefont {{Torrontegui}}\ \emph {et~al.}(2013)\citenamefont
  {{Torrontegui}}, \citenamefont {{Ib{\'a}{\~n}ez}}, \citenamefont
  {{Mart{\'\i}nez-Garaot}}, \citenamefont {{Modugno}}, \citenamefont {{del
  Campo}}, \citenamefont {{Gu{\'e}ry-Odelin}}, \citenamefont {{Ruschhaupt}},
  \citenamefont {{Chen}},\ and\ \citenamefont {{Muga}}}]{Torrontegui_2015}%
  \BibitemOpen
  \bibfield  {author} {\bibinfo {author} {\bibfnamefont {E.}~\bibnamefont
  {{Torrontegui}}}, \bibinfo {author} {\bibfnamefont {S.}~\bibnamefont
  {{Ib{\'a}{\~n}ez}}}, \bibinfo {author} {\bibfnamefont {S.}~\bibnamefont
  {{Mart{\'\i}nez-Garaot}}}, \bibinfo {author} {\bibfnamefont {M.}~\bibnamefont
  {{Modugno}}}, \bibinfo {author} {\bibfnamefont {A.}~\bibnamefont {{del
  Campo}}}, \bibinfo {author} {\bibfnamefont {D.}~\bibnamefont
  {{Gu{\'e}ry-Odelin}}}, \bibinfo {author} {\bibfnamefont {A.}~\bibnamefont
  {{Ruschhaupt}}}, \bibinfo {author} {\bibfnamefont {X.}~\bibnamefont
  {{Chen}}},\ and\ \bibinfo {author} {\bibfnamefont {J.~G.}\ \bibnamefont
  {{Muga}}},\ }\href {https://doi.org/10.1016/B978-0-12-408090-4.00002-5}
  {\bibfield  {journal} {\bibinfo  {journal} {Advances in Atomic Molecular and
  Optical Physics}\ }\textbf {\bibinfo {volume} {62}},\ \bibinfo {pages} {117}
  (\bibinfo {year} {2013})}\BibitemShut {NoStop}%
\bibitem [{\citenamefont {Schaff}\ \emph {et~al.}(2011)\citenamefont {Schaff},
  \citenamefont {Song}, \citenamefont {Capuzzi}, \citenamefont {Vignolo},\ and\
  \citenamefont {Labeyrie}}]{Schaff_2011}%
  \BibitemOpen
  \bibfield  {author} {\bibinfo {author} {\bibfnamefont {J.-F.}\ \bibnamefont
  {Schaff}}, \bibinfo {author} {\bibfnamefont {X.-L.}\ \bibnamefont {Song}},
  \bibinfo {author} {\bibfnamefont {P.}~\bibnamefont {Capuzzi}}, \bibinfo
  {author} {\bibfnamefont {P.}~\bibnamefont {Vignolo}},\ and\ \bibinfo {author}
  {\bibfnamefont {G.}~\bibnamefont {Labeyrie}},\ }\href
  {https://doi.org/10.1209/0295-5075/93/23001} {\bibfield  {journal} {\bibinfo
  {journal} {{EPL} (Europhysics Letters)}\ }\textbf {\bibinfo {volume} {93}},\
  \bibinfo {pages} {23001} (\bibinfo {year} {2011})}\BibitemShut {NoStop}%
\bibitem [{\citenamefont {An}\ \emph {et~al.}(2016)\citenamefont {An},
  \citenamefont {Lv}, \citenamefont {del Campo},\ and\ \citenamefont
  {Kim}}]{An_2016}%
  \BibitemOpen
  \bibfield  {author} {\bibinfo {author} {\bibfnamefont {S.}~\bibnamefont
  {An}}, \bibinfo {author} {\bibfnamefont {D.}~\bibnamefont {Lv}}, \bibinfo
  {author} {\bibfnamefont {A.}~\bibnamefont {del Campo}},\ and\ \bibinfo
  {author} {\bibfnamefont {K.}~\bibnamefont {Kim}},\ }\href
  {https://doi.org/10.1038/ncomms12999} {\bibfield  {journal} {\bibinfo
  {journal} {Nature Communications}\ }\textbf {\bibinfo {volume} {7}},\
  \bibinfo {pages} {12999} (\bibinfo {year} {2016})}\BibitemShut {NoStop}%
\bibitem [{\citenamefont {Mart{\'i}nez}\ \emph {et~al.}(2016)\citenamefont
  {Mart{\'i}nez}, \citenamefont {Petrosyan}, \citenamefont {Gu{\'e}ry-Odelin},
  \citenamefont {Trizac},\ and\ \citenamefont {Ciliberto}}]{Martinez_2016}%
  \BibitemOpen
  \bibfield  {author} {\bibinfo {author} {\bibfnamefont {I.~A.}\ \bibnamefont
  {Mart{\'i}nez}}, \bibinfo {author} {\bibfnamefont {A.}~\bibnamefont
  {Petrosyan}}, \bibinfo {author} {\bibfnamefont {D.}~\bibnamefont
  {Gu{\'e}ry-Odelin}}, \bibinfo {author} {\bibfnamefont {E.}~\bibnamefont
  {Trizac}},\ and\ \bibinfo {author} {\bibfnamefont {S.}~\bibnamefont
  {Ciliberto}},\ }\href {https://doi.org/10.1038/nphys3758} {\bibfield
  {journal} {\bibinfo  {journal} {Nature Physics}\ }\textbf {\bibinfo {volume}
  {12}},\ \bibinfo {pages} {843} (\bibinfo {year} {2016})}\BibitemShut
  {NoStop}%
\bibitem [{\citenamefont {Dann}\ \emph {et~al.}(2019)\citenamefont {Dann},
  \citenamefont {Tobalina},\ and\ \citenamefont {Kosloff}}]{Dann_2019}%
  \BibitemOpen
  \bibfield  {author} {\bibinfo {author} {\bibfnamefont {R.}~\bibnamefont
  {Dann}}, \bibinfo {author} {\bibfnamefont {A.}~\bibnamefont {Tobalina}},\
  and\ \bibinfo {author} {\bibfnamefont {R.}~\bibnamefont {Kosloff}},\ }\href
  {https://doi.org/10.1103/PhysRevLett.122.250402} {\bibfield  {journal}
  {\bibinfo  {journal} {Phys. Rev. Lett.}\ }\textbf {\bibinfo {volume} {122}},\
  \bibinfo {pages} {250402} (\bibinfo {year} {2019})}\BibitemShut {NoStop}%
\bibitem [{\citenamefont {Villazon}\ \emph {et~al.}(2019)\citenamefont
  {Villazon}, \citenamefont {Polkovnikov},\ and\ \citenamefont
  {Chandran}}]{Villazon_2019}%
  \BibitemOpen
  \bibfield  {author} {\bibinfo {author} {\bibfnamefont {T.}~\bibnamefont
  {Villazon}}, \bibinfo {author} {\bibfnamefont {A.}~\bibnamefont
  {Polkovnikov}},\ and\ \bibinfo {author} {\bibfnamefont {A.}~\bibnamefont
  {Chandran}},\ }\href {https://doi.org/10.1103/PhysRevA.100.012126} {\bibfield
   {journal} {\bibinfo  {journal} {Phys. Rev. A}\ }\textbf {\bibinfo {volume}
  {100}},\ \bibinfo {pages} {012126} (\bibinfo {year} {2019})}\BibitemShut
  {NoStop}%
\bibitem [{\citenamefont {Anderegg}\ \emph {et~al.}(2019)\citenamefont
  {Anderegg}, \citenamefont {Cheuk}, \citenamefont {Bao}, \citenamefont
  {Burchesky}, \citenamefont {Ketterle}, \citenamefont {Ni},\ and\
  \citenamefont {Doyle}}]{Anderegg_2019}%
  \BibitemOpen
  \bibfield  {author} {\bibinfo {author} {\bibfnamefont {L.}~\bibnamefont
  {Anderegg}}, \bibinfo {author} {\bibfnamefont {L.~W.}\ \bibnamefont {Cheuk}},
  \bibinfo {author} {\bibfnamefont {Y.}~\bibnamefont {Bao}}, \bibinfo {author}
  {\bibfnamefont {S.}~\bibnamefont {Burchesky}}, \bibinfo {author}
  {\bibfnamefont {W.}~\bibnamefont {Ketterle}}, \bibinfo {author}
  {\bibfnamefont {K.-K.}\ \bibnamefont {Ni}},\ and\ \bibinfo {author}
  {\bibfnamefont {J.~M.}\ \bibnamefont {Doyle}},\ }\href
  {https://doi.org/10.1126/science.aax1265} {\bibfield  {journal} {\bibinfo
  {journal} {Science}\ }\textbf {\bibinfo {volume} {365}},\ \bibinfo {pages}
  {1156} (\bibinfo {year} {2019})}\BibitemShut {NoStop}%
\bibitem [{\citenamefont {Omran}\ \emph {et~al.}(2019)\citenamefont {Omran},
  \citenamefont {Levine}, \citenamefont {Keesling}, \citenamefont {Semeghini},
  \citenamefont {Wang}, \citenamefont {Ebadi}, \citenamefont {Bernien},
  \citenamefont {Zibrov}, \citenamefont {Pichler}, \citenamefont {Choi},
  \citenamefont {Cui}, \citenamefont {Rossignolo}, \citenamefont {Rembold},
  \citenamefont {Montangero}, \citenamefont {Calarco}, \citenamefont {Endres},
  \citenamefont {Greiner}, \citenamefont {Vuleti{\'c}},\ and\ \citenamefont
  {Lukin}}]{Omran_2019}%
  \BibitemOpen
  \bibfield  {author} {\bibinfo {author} {\bibfnamefont {A.}~\bibnamefont
  {Omran}}, \bibinfo {author} {\bibfnamefont {H.}~\bibnamefont {Levine}},
  \bibinfo {author} {\bibfnamefont {A.}~\bibnamefont {Keesling}}, \bibinfo
  {author} {\bibfnamefont {G.}~\bibnamefont {Semeghini}}, \bibinfo {author}
  {\bibfnamefont {T.~T.}\ \bibnamefont {Wang}}, \bibinfo {author}
  {\bibfnamefont {S.}~\bibnamefont {Ebadi}}, \bibinfo {author} {\bibfnamefont
  {H.}~\bibnamefont {Bernien}}, \bibinfo {author} {\bibfnamefont {A.~S.}\
  \bibnamefont {Zibrov}}, \bibinfo {author} {\bibfnamefont {H.}~\bibnamefont
  {Pichler}}, \bibinfo {author} {\bibfnamefont {S.}~\bibnamefont {Choi}},
  \bibinfo {author} {\bibfnamefont {J.}~\bibnamefont {Cui}}, \bibinfo {author}
  {\bibfnamefont {M.}~\bibnamefont {Rossignolo}}, \bibinfo {author}
  {\bibfnamefont {P.}~\bibnamefont {Rembold}}, \bibinfo {author} {\bibfnamefont
  {S.}~\bibnamefont {Montangero}}, \bibinfo {author} {\bibfnamefont
  {T.}~\bibnamefont {Calarco}}, \bibinfo {author} {\bibfnamefont
  {M.}~\bibnamefont {Endres}}, \bibinfo {author} {\bibfnamefont
  {M.}~\bibnamefont {Greiner}}, \bibinfo {author} {\bibfnamefont
  {V.}~\bibnamefont {Vuleti{\'c}}},\ and\ \bibinfo {author} {\bibfnamefont
  {M.~D.}\ \bibnamefont {Lukin}},\ }\href
  {https://doi.org/10.1126/science.aax9743} {\bibfield  {journal} {\bibinfo
  {journal} {Science}\ }\textbf {\bibinfo {volume} {365}},\ \bibinfo {pages}
  {570} (\bibinfo {year} {2019})}\BibitemShut {NoStop}%
\bibitem [{\citenamefont {Landau}\ and\ \citenamefont
  {Lifshitz}(1965)}]{Landau_1965}%
  \BibitemOpen
  \bibfield  {author} {\bibinfo {author} {\bibfnamefont {L.~D.}\ \bibnamefont
  {Landau}}\ and\ \bibinfo {author} {\bibfnamefont {E.~M.}\ \bibnamefont
  {Lifshitz}},\ }\href@noop {} {\emph {\bibinfo {title} {Quantum Mechanics}}}\
  (\bibinfo  {publisher} {Pergamon Press},\ \bibinfo {year} {1965})\BibitemShut
  {NoStop}%
\bibitem [{\citenamefont {Zener}\ and\ \citenamefont
  {Fowler}(1932)}]{Zener_1932}%
  \BibitemOpen
  \bibfield  {author} {\bibinfo {author} {\bibfnamefont {C.}~\bibnamefont
  {Zener}}\ and\ \bibinfo {author} {\bibfnamefont {R.~H.}\ \bibnamefont
  {Fowler}},\ }\href {https://doi.org/10.1098/rspa.1932.0165} {\bibfield
  {journal} {\bibinfo  {journal} {Proceedings of the Royal Society of London.
  Series A, Containing Papers of a Mathematical and Physical Character}\
  }\textbf {\bibinfo {volume} {137}},\ \bibinfo {pages} {696} (\bibinfo {year}
  {1932})}\BibitemShut {NoStop}%
\bibitem [{\citenamefont {Takhtadzhan}\ and\ \citenamefont
  {Faddeev}(1979)}]{Takhtadzhan_1979}%
  \BibitemOpen
  \bibfield  {author} {\bibinfo {author} {\bibfnamefont {L.~A.}\ \bibnamefont
  {Takhtadzhan}}\ and\ \bibinfo {author} {\bibfnamefont {L.~D.}\ \bibnamefont
  {Faddeev}},\ }\href {https://doi.org/10.1070/rm1979v034n05abeh003909}
  {\bibfield  {journal} {\bibinfo  {journal} {Russian Mathematical Surveys}\
  }\textbf {\bibinfo {volume} {34}},\ \bibinfo {pages} {11} (\bibinfo {year}
  {1979})}\BibitemShut {NoStop}%
\bibitem [{\citenamefont {Deffner}\ and\ \citenamefont
  {Lutz}(2010)}]{Deffner_2010}%
  \BibitemOpen
  \bibfield  {author} {\bibinfo {author} {\bibfnamefont {S.}~\bibnamefont
  {Deffner}}\ and\ \bibinfo {author} {\bibfnamefont {E.}~\bibnamefont {Lutz}},\
  }\href {https://doi.org/10.1103/PhysRevLett.105.170402} {\bibfield  {journal}
  {\bibinfo  {journal} {Phys. Rev. Lett.}\ }\textbf {\bibinfo {volume} {105}},\
  \bibinfo {pages} {170402} (\bibinfo {year} {2010})}\BibitemShut {NoStop}%
\bibitem [{\citenamefont {Baumgratz}\ \emph {et~al.}(2014)\citenamefont
  {Baumgratz}, \citenamefont {Cramer},\ and\ \citenamefont
  {Plenio}}]{Baumgratz_2014}%
  \BibitemOpen
  \bibfield  {author} {\bibinfo {author} {\bibfnamefont {T.}~\bibnamefont
  {Baumgratz}}, \bibinfo {author} {\bibfnamefont {M.}~\bibnamefont {Cramer}},\
  and\ \bibinfo {author} {\bibfnamefont {M.~B.}\ \bibnamefont {Plenio}},\
  }\href {https://doi.org/10.1103/PhysRevLett.113.140401} {\bibfield  {journal}
  {\bibinfo  {journal} {Phys. Rev. Lett.}\ }\textbf {\bibinfo {volume} {113}},\
  \bibinfo {pages} {140401} (\bibinfo {year} {2014})}\BibitemShut {NoStop}%
\bibitem [{\citenamefont {Vacanti}\ \emph {et~al.}(2014)\citenamefont
  {Vacanti}, \citenamefont {Fazio}, \citenamefont {Montangero}, \citenamefont
  {Palma}, \citenamefont {Paternostro},\ and\ \citenamefont
  {Vedral}}]{Vacanti_2014}%
  \BibitemOpen
  \bibfield  {author} {\bibinfo {author} {\bibfnamefont {G.}~\bibnamefont
  {Vacanti}}, \bibinfo {author} {\bibfnamefont {R.}~\bibnamefont {Fazio}},
  \bibinfo {author} {\bibfnamefont {S.}~\bibnamefont {Montangero}}, \bibinfo
  {author} {\bibfnamefont {G.~M.}\ \bibnamefont {Palma}}, \bibinfo {author}
  {\bibfnamefont {M.}~\bibnamefont {Paternostro}},\ and\ \bibinfo {author}
  {\bibfnamefont {V.}~\bibnamefont {Vedral}},\ }\href
  {https://doi.org/10.1088/1367-2630/16/5/053017} {\bibfield  {journal}
  {\bibinfo  {journal} {New Journal of Physics}\ }\textbf {\bibinfo {volume}
  {16}},\ \bibinfo {pages} {053017} (\bibinfo {year} {2014})}\BibitemShut
  {NoStop}%
\end{thebibliography}%

\onecolumngrid
\appendix
\section{Derivation of the counterdiabatic term}\label{app:derivationHtqd}
One derivation of $\hH_{tqd}(t)$ in discrete-variable quantum systems is given as follows \cite{Vacanti_2014}:

Suppose the system Hamiltonian is $\hH_s(t)$ and its instantaneous eigenstates are $\{\ket{n(t)}\}$. By choosing a time-independent basis $\{\ket{n(0)}\}$, one can define the similarity transformation
\begin{equation}
\hU(t)=\sum_n \ket{n(t)}\bra{n(0)}.
\end{equation}
Multiplying $\hU^{-1}$ on both sides of the time-dependent Schr\"odinger equation (TDSE), we express the equation in basis of its instantaneous eigenstates  (we drop the time notation from here),
\begin{equation}
\hU^{-1} i\hbar \dot{\ket{n}} = \hU^{-1} \hH_s \ket{n},
\end{equation}
where $\dot{\ket{n}}$ denotes the time derivative of $\ket{n}$. Define the Hamiltonian and the state in instantaneous basis, such that $\hH_d=\hU^{-1}\hH_s\hU$ and $\ket{n}_d=\hU^{-1}\ket{n}$, the above equation becomes
\begin{equation}\label{equ:tqdstep1}
i\hbar \hU^{-1}\dot{\ket{n}} = \hU^{-1} \hH_s \hU \hU^{-1} \ket{n} =\hH_d\ket{n}_d.
\end{equation}
where $\hU\hU^{-1}=\hI$. Taking the time derivative of $\ket{n}_d$, one gets $\dot{\ket{n}}_d=\dot{U}^{-1}\ket{n}+U^{-1}\dot{\ket{n}}$. Apply this to \eq{equ:tqdstep1}, we get
\begin{equation}
i\hbar\dot{\ket{n}}_d = \left[ \hH_d + i\hbar \dot{\hU}^{-1} \hU \right] \ket{n}_d
\end{equation}
Here $\hH_d$ is a diagonal matrix but $i\hbar\dot{\hU}^{-1}\hU$ is usually not. We can decompose $i\hbar\dot{\hU}^{-1}\hU$ as the sum of a diagonal term $\hH'_d$ and a non-diagonal term $\hH'_{nd}$, such that
\begin{equation}\label{equ:tqdstep3}
i\hbar\dot{\ket{n}}_d = \left[ \hH_d + \hH'_d + \hH'_{nd} \right] \ket{n}_d.
\end{equation}
To avoid transitions, we wish to only keep the diagonal terms $\hH_d$ and $\hH'_d$, and remove the non-diagonal term $\hH'_{nd}$. This can be done by defining a counterdiabatic Hamiltonian 
\begin{equation}
\hH_{tqd} = - \hU \hH'_{nd} \hU^{-1},
\end{equation}
such that adding this additional term to the original system Hamiltonian cancels out the non-diagonal term $\hH'_{nd}$ in \eq{equ:tqdstep3}, and changing the remaining TDSE into
\begin{equation}
i\hbar\dot{\ket{n}}_d = \left[ \hH_d + \hH'_d  \right] \ket{n}_d.
\end{equation}

\end{document}